\documentclass[preprints,article,accept,moreauthors,pdftex]{mdpi} 
\usepackage{xcolor}
\usepackage{siunitx}
\usepackage{amssymb}
\usepackage{amsmath}
\usepackage{amsthm}
\usepackage{comment}

\firstpage{1} 
\pubvolume{}
\issuenum{}
\articlenumber{}
\pubyear{}
\copyrightyear{}
\history{}
\newcommand{\SupC}{{I}}

\newcommand{\ElEn}{{\cal E}}
\newcommand{\ph}{{\varphi}}

\newcommand{\DoSF}{{{\cal N}_0}}

\newcommand{\artanh}{\mathop{\mathrm{arctanh}}}
\newcommand{\sech}{{\mathop{\mathrm{sech}}}}
\newcommand{\en}{\varepsilon}
\newcommand{\dd}{{\rm d}}

\newcommand{\RV}{$v_R\, $}
\newcommand{\LV}{$v_L\, $}
\newcommand{\COP}{{\rm COP}}

\newcommand{\TRes}{\overline{T}}

\Title{Thermodynamics of a phase-driven proximity Josephson junction}

\Author{Francesco Vischi$^{1,2,}$*\orcidA{}, Matteo Carrega$^1$, Alessandro Braggio$^{1}$, Pauli Virtanen$^{3}$ and Francesco Giazotto$^{1}$}

\AuthorNames{Francesco Vischi, Matteo Carrega, Alessandro Braggio, Francesco Giazotto$^{1}$}

\address{%
$^{1}$ \quad NEST, Istituto Nanoscienze-CNR and Scuola Normale Superiore, Piazza S. Silvestro 12, I-56127 Pisa, Italy\\
$^{2}$ \quad Dipartimento di Fisica ”E. Fermi”, Università di Pisa, Largo Bruno Pontecorvo 3, I-56127 Pisa, Italy\\
$^{3}$ \quad University of Jyv\"{a}skyl\"{a}, Department of Physics and Nanoscience Center, P.O. Box 35 (YFL), FI-40014, Finland
}

\corres{Correspondence: francesco.vischi@df.unipi.it}

\abstract{
	\textls[-10]{We study the thermodynamic properties of a superconductor/normal metal/superconductor Josephson junction {in the short limit}. Owing to the proximity effect, such a junction constitutes a thermodynamic system where {phase difference}, supercurrent, temperature and entropy are thermodynamical variables connected by equations of state.  These allow conceiving quasi-static processes that we characterize in terms of heat and work exchanged. Finally, we combine such processes to construct a Josephson-based Otto and Stirling cycles. We study the related performance in both engine and refrigerator operating mode.}
}

\begin{document}
	\section{Introduction}
	{Thermodynamic concepts have been recently considered at the nanoscale, conceiving and realizing systems where quantum
		coherent properties are mirrored in thermodynamic quantities at {mesoscopic} level 
		\cite{streltsov2017, vinjanampathy2016, pekola2015, kosloff2013, vinjanampathy2016, kosloff2017, stefano2008, goold2016,guiducci2019, guiducci2019PRB,  giazotto2006PRL,fornieri2017, hofmann2016, campisi2011, esposito2009}
		. Furthermore, one of the most impressive examples of quantum features reflected in macroscopic systems is represented by superconductivity, where quantum coherence is manifested at a {mesoscopic} scale. Therefore superconducting systems are interesting platform where investigating the interplay between thermodynamic concepts and quantum coherences.}
	
	Superconducting hybrid systems, i.e.,  constituted of superconducting parts in electric contact with normal (non-superconducting) parts, are in practice coherent electron systems with striking thermodynamic equilibrium/transport properties, resulting in a wide variety of applicative devices: low-temperature sensitive thermometers \cite{wang2018,zgirski2018,gasparinetti2015,giazotto2006,giazotto2015},
	sensitive detectors \cite{govenius2016,lee2019, giazotto2008, guarcello2019,vischi2019,solinas2018,virtanen2018,koppens2014,mitrofanov2017,giordano2018,giazotto2008,mckitterick2015,mckitterick2016,du2008},  heat valves \cite{sothmann2017,yang2019,goffman2017,dutta2017,li2012,joulain2016, ronzani2018,sivre2017,strambini2014,giazotto2012,bours2019},
	caloritronics (heat computing) \cite{wang2007,li2012, paolucci2018logic, fornieri2017,hwang2019,bauer2019, timossi2018}, solid-state micro-refrigerators \cite{muhonen2012,leivo1996,courtois2016,nguyen2016,giazotto2006,solinas2016,koppinen2009, sanchez2019}, solid-state
	quantum machines \cite{karimi2016,marchegiani2016,vischi2018,haack2019,manikandan2019,carrega2019}, thermoelectric generators \cite{benenti2017, heikkila2019, marchegiani2018,sanchez2016,hussein2019,marchegiani2019}.
	
	In this paper, we review the equilibrium thermodynamic properties of a hybrid system based on a Superconductor/Normal metal/Superconductor {(SNS) Josephson Junction} in the diffusive limit. The~behavior of such a system is ruled by the proximity effect, which consists in a set of physical phenomena owing to the propagation of the superconducting electron correlations in the normal \mbox{metal \cite{pannetier2000,courtois1999,likharev1979}}. In~particular, guided by a matter of thermodynamic consistency, we discuss a relation between the electronic and thermal properties of the proximized system. From this relation, we~develop a basic investigation of
	the thermodynamic properties of such a system. These results are then exploited to investigate quasi-static processes and thermodynamic cycles. {We focus within a semi-classical regime of Josephson coupling, i.e.,  we neglect non-commutativity between the phase and the number of pairs, as usually done in the thermodynamic limit.}
	
	{We remark} {that, besides the system studied in this paper, many equilibrium thermodynamic properties have been investigated in different conditions, theoretically and experimentally: thermodynamics of rings interrupted by insulating Josephson Junction \cite{debruynouboter1988,debruynouboter1989, vandenbrink1997a,vandenbrink1997b,vleeming1997}, heat capacity in SN systems \cite{lechevet1972,manuel1976,zaitlin1982}, free energy in hybrid SN systems due to boundary effects with approaches different to the quasi-classical theory \cite{kobes1988,kosztin1998,eilenberger1975}.}
	
	\textls[-10]{The paper is organized as follows. Section \ref{sec:Thermodynamics} describes the proximized system under study and introduces its thermodynamics, giving also
		an insight into the underlying microscopical mechanism. Section \ref{sec:ThermodynamicProcesses} studies the thermodynamic processes. Hence, these are combined
		in Section \ref{sec:ThermodynamicsCycles} to investigate two different thermodynamic cycles. 
		Finally, Section \ref{sec:Discussion} summarizes and discusses the main findings. For completeness, Appendix \ref{sec:Limits} discusses the thermodynamics of a Josephson junction close to the critical temperature.}
	
	\section{Thermodynamics of Hybrid Systems}
	\label{sec:Thermodynamics}
	\subsection{Model}
	\label{sec:Model}

\begin{figure} [t]
		\centering
		\includegraphics[width=0.95\textwidth]{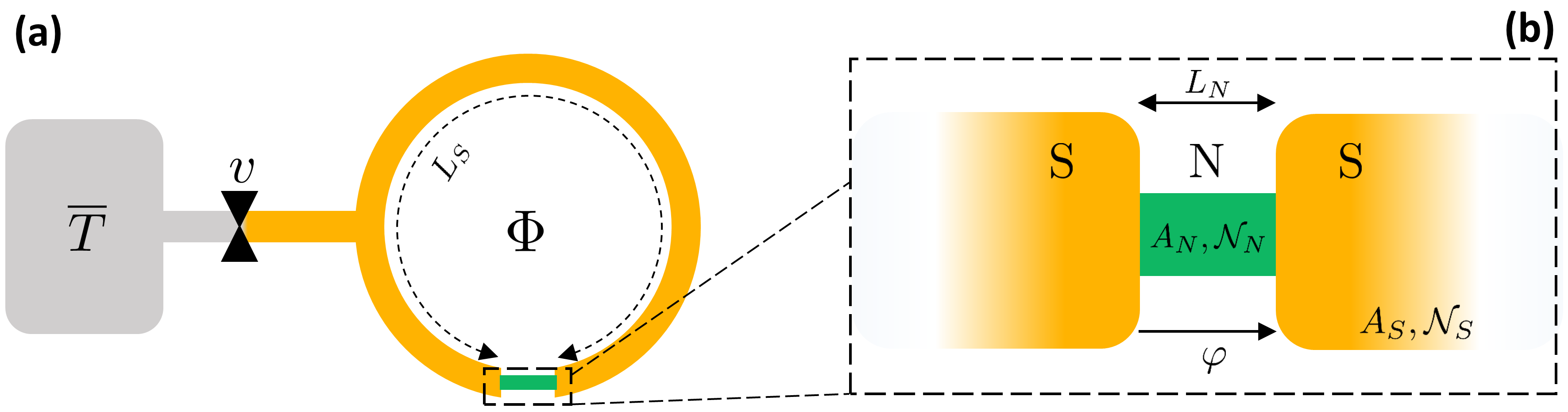}
		\caption{({\bf a}) Sketch of the SNS proximized system. It consists of superconducting ring, $L_S$ long, pierced by a magnetic flux $\Phi$. The ring is interrupted by a normal metal weak link. The electron system of the whole device is thermally and electrically isolated and at temperature $T$. The system is connected to a thermal reservoir at temperature $\overline T$ through a heat valve $v$. ({\bf b}) Magnification of the SNS junction. The~normal metal weak, $L_N$ long, is in clean electric contact with the superconducting leads. $A_j,{\cal N}_j$ are respectively the cross-section and the DoS at Fermi energy of the $j=N$ or $S$ metal. The phase drop $\ph$ of the superconducting order parameter takes place across the junction. 
		}
		\label{fig:Sketch}
	\end{figure}

	We consider a system as sketched in Figure \ref{fig:Sketch}, constituted by a superconducting ring interrupted by a Superconductor/Normal metal/Superconductor
	(SNS) proximity Josephson Junction. The~superconductor gap depending on temperature $T$ is $\Delta(T)$ and reaches $\Delta_0$ at $T=0$. The~critical temperature is $T_c$.  The phase difference $\ph$ of the superconducting order parameter across the junction is ruled by the magnetic flux threading the
	ring, owing to the fluxoid quantization relation $\ph=2\pi \Phi/\Phi_0$, where $\Phi_0=h/2e\approx\SI
	{2E-15}{\weber}$ is the flux quantum. 
	
	\textls[-10]{We assume that the system is thermally and electrically isolated and at a homogeneous  temperature $T$, neglecting thermal gradients. We consider only heat exchange with a reservoir at temperature $\TRes$ through the respective heat channel
		connected by a heat valve $v$ \cite{sothmann2017,yang2019,goffman2017,dutta2017,joulain2016,strambini2014,li2012,sivre2017,ronzani2018,giazotto2012}, as drawn in Figure \ref{fig:Sketch}.}
	
	The junction, magnified in Figure \ref{fig:Sketch}b, consists in the two S leads in electric contact with an N weak link. The superconductor has a critical
	temperature $T_c$ and BCS gap at zero temperature $\Delta_0$. The N weak link and the S leads have respectively cross-sections $A_N$ and $A_S$, conductivities
	$\sigma_N$ and $\sigma_S$, Density of States (DoS) per spin at the Fermi level ${\cal N}_N$ and ${\cal N}_S$. The length of the weak link is $L_N$, resulting in a resistance $R_N=L_N/A_N\sigma_N$. The length of the superconducting ring is $L_S$. The whole proximized system (ring+junction) volume is $V$. 
	
	We make the following assumptions about the junction, in order to make simple analytical predictions within the Kulik-Omel'yanchuk (KO) treatment \cite{kulik1975,golubov2004,virtanen2017SNS}.
	We consider diffusive charge transport with diffusivity $D$ for both the S and N parts.
	This requires that the weak link is longer than the mean free path $L_{\rm mfp}$:  $L_N\gg L_{\rm mfp}$. The KO treatment holds when the whole junction can be treated in a quasi-1 dimensional approximation, i.e.,  when  $A_S,A_N \ll \xi^2$. The diffusivity defines the coherence length $\xi =
	\sqrt{\hbar D/\Delta_0 }$ \cite{hammer2007}. Moreover, we consider a short constriction weak link respect to the superconducting leads. Quantitatively, using the parameters $l=L_N/\xi$ and
	$a=\sigma _S A_S/\sigma_N A_N$, we consider a short junction with $l \lesssim 1$  and a constriction with $al \gg 1$.
	
	The typical values for this kind of system are the following. The ring can be made of aluminium, with $\Delta_0 \approx 180\,$eV, corresponding to $T_c\approx 1.2$K \cite{cochran1958,giaever1961, langenberg1966}. The coherence length for hybrid Al-based devices is about $\xi \approx 150\,$nm \cite{dambrosio2015, giazotto2010,ronzani2014}. In the following, we set the Boltzmann constant to $k_B=1$, implying that the temperatures have a physical dimension of energy while entropy and specific heat are dimensionless.
	
	\subsection{Hybrid Junction as Thermodynamic System}
	\label{sec:JunctionAsTDSystem}
	Before investigating the thermodynamic behavior of our system in detail, we discuss about the thermodynamic consistency under a general point of view that is valid for any Josephson Junction (JJ). In particular, we focus on the relation between the current transport and the junction entropy.
	
	In a JJ, the Current Phase Relation (CPR) describes the  dissipationless supercurrent $I(\ph,T)$ flowing across it as function of the phase
	difference $\ph$ and temperature $T$ \cite{golubov2004,likharev1979}. The precise form of the CPR depends on the geometry and on the materials of the junction, and can be
	calculated from the free energy as
	\begin{equation}
	\frac{\hbar}{2e} I(\ph,T) = \frac{\partial F(\ph,T)}{\partial \ph} \, \, 
	\end{equation}
	where $F(\ph,T)$ has to be calculated within quantum statistical methods as a function of the state variables $(\ph,T)$. The
	CPR constitutes an equation of state connecting $I,\ph$ and $T$. Another equation of state is given by the entropy $S(\ph,T)$ as a function of phase difference and temperature
	\begin{equation}
	S(\ph,T) = -\frac{\partial F(\ph,T)}{\partial T} \, \, .
	\end{equation}
	
	The entropy and the CPR are necessarily linked by thermodynamic consistency. Indeed the two cross derivatives of $F$ are identical, i.e.,  $\partial_\ph
	\partial_T F = \partial_T \partial_\ph F$, owing to the Schwarz theorem. Hence, the following Maxwell relation is universally valid
	\begin{equation}
	-\frac{\partial S(\ph,T)}{\partial \ph} = \frac{\hbar}{2e}\frac{\partial I(\ph,T)}{\partial T} \, \, .
	\label{eq:MaxwellRelation}
	\end{equation}
	Using this equation, the entropy of the JJ can be expressed as 
	\begin{equation}
	S(\ph,T) = S_0(T) + \delta S (\ph,T) \, \, 
	\label{eq:Stot}
	\end{equation}
	where $S_0(T)$ is the entropy at $\ph=0$ and $\delta S(\ph,T)$ is the phase-dependent entropy variation
	\begin{align}
	&    \label{eq:deltaS}  \delta S (\ph,T) = -\frac{\partial}{\partial T} \ElEn (\ph,T)\\
	& \ElEn(\ph,T) = \frac{e R_0}{2\pi} \int _0 ^\ph I(\ph',T) \dd \ph'  \, \, 
	\label{eq:ElEnDefinition}
	\end{align}
	where $\ElEn(\ph,T)$ is the Josephson energy stored in the junction at a given temperature $T$, $R_0 = h/2e^2 \approx \SI{12.9}{\kilo\ohm}$ is the inverse
	of the quantum of conductance. We note that the prefactor in Equation (\ref{eq:ElEnDefinition}) is usually expressed as $\Phi_0/2\pi$. We chose the form $e R_0/2\pi$ to allow an easier comparison with the junction resistance $R_N$. 
	
	The entropy $S_0(T)$ at $\ph=0$ cannot be determined from the knowledge of the CPR. Indeed, any function $S_0(T)$ of the temperature is dropped by the phase
	derivative in Equation (\ref{eq:MaxwellRelation}), hence satisfying the Maxwell equation. The physical solution of $S_0(T)$ can be found within a microscopic model that we show in the next subsection.
	
	\subsection{Proximity Induced Minigap}
	\label{sec:ProximityMechanism}
	In this subsection, we give an insight into the microscopic mechanism which determines the entropy in a hybrid junction. In particular, we show that the entropy
	dependence on temperature and phase is related to the presence of an induced phase-dependent minigap in the quasi-particle Density of States (DoS).
	In a hybrid NS, correlated electrons propagate from the superconductor into the normal metal, strongly modifying the
	properties of the latter with a set of phenomena called generically under the name of   proximity effect \cite{pannetier2000,courtois1999,taddei2005,kulik2012,belzig1999,nazarov1994,
		nazarov1999,heedt2017}. Among all possible consequences dictated by the proximity effect, here we focus on the induced mini-gap in the quasi-particle Density of States, it being responsible for the phase and temperature dependence of the entropy $S$ in an SNS junction.
	
	Let us consider the N weak link in an SNS junction. When not proximized, the weak link DoS is homogeneous and approximately constant at its Fermi level
	value ${\cal N}_N$ in the energy range of interest of few $\Delta_0$ around the Fermi energy. Instead, when proximized by the superconducting leads, the DoS is no more constant
	neither on energy nor on position, but is given by ${\cal N}_N N(\mathbf{r},\en,\ph)$, where $N(\mathbf{r},\en,\ph)$ is the normalized local DoS \cite{hammer2007,zhou1998,belzig1999,bergeret2008}
	that is a function of the position $\mathbf{r}$, energy $\en$ and the phase difference $\ph$.
	
	One way to calculate the normalized local DoS is provided by the quasi-classical theory of superconductivity \cite{hammer2007,zhou1998, belzig1999, bergeret2008, nazarov1994, nazarov1999,vischi2017}. Qualitatively, a result of this theory is that the normalized local DoS is characterized by an induced gap in the N weak link, whose amplitude $\tilde \Delta$ is smaller than the S bulk gap $\Delta(T)$. For this reason, $\tilde
	\Delta$ is dubbed induced minigap. This induced minigap has the following properties \cite{zhou1998,hammer2007}: its width $\tilde \Delta $ at $\ph=0$ depends
	on the weak link length $L_N$ and reaches $\tilde \Delta\to \Delta(T)$ when  $L_N$ is well below the coherence length $\xi$. Moreover, $\tilde{\Delta}$
	depends on the phase $\ph$ through a function that is even and $2\pi$ periodic. The minigap is fully open at $\ph=0$ and shrinks till closure at $\ph=\pi$.
	An~analytical solution of the local normalized DoS is available for diffusive short junctions with rigid boundary conditions \cite{heikkila2002,giazotto2011},
	yielding that $\tilde{\Delta} = \Delta(T) |\cos (\ph/2)|$. The proximity induced gap and its interesting properties have been observed experimentally by tunneling
	experiments \cite{lesueur2008,giazotto2010,dambrosio2015}.
	
	An important feature of this microscopic proximity DoS modification is that it does not take place just in the N weak link, but also affects the S leads as well. The anti-proximization operated by the N weak link on the S leads is called {\it inverse proximity effect} and plays the role of a crucial correction in short junctions,
	since it gives an important contribution to the total entropy dependence on the junction phase \cite{virtanen2017SNS}.

\begin{figure} [t]
		\centering
		\includegraphics[width=0.95\textwidth]{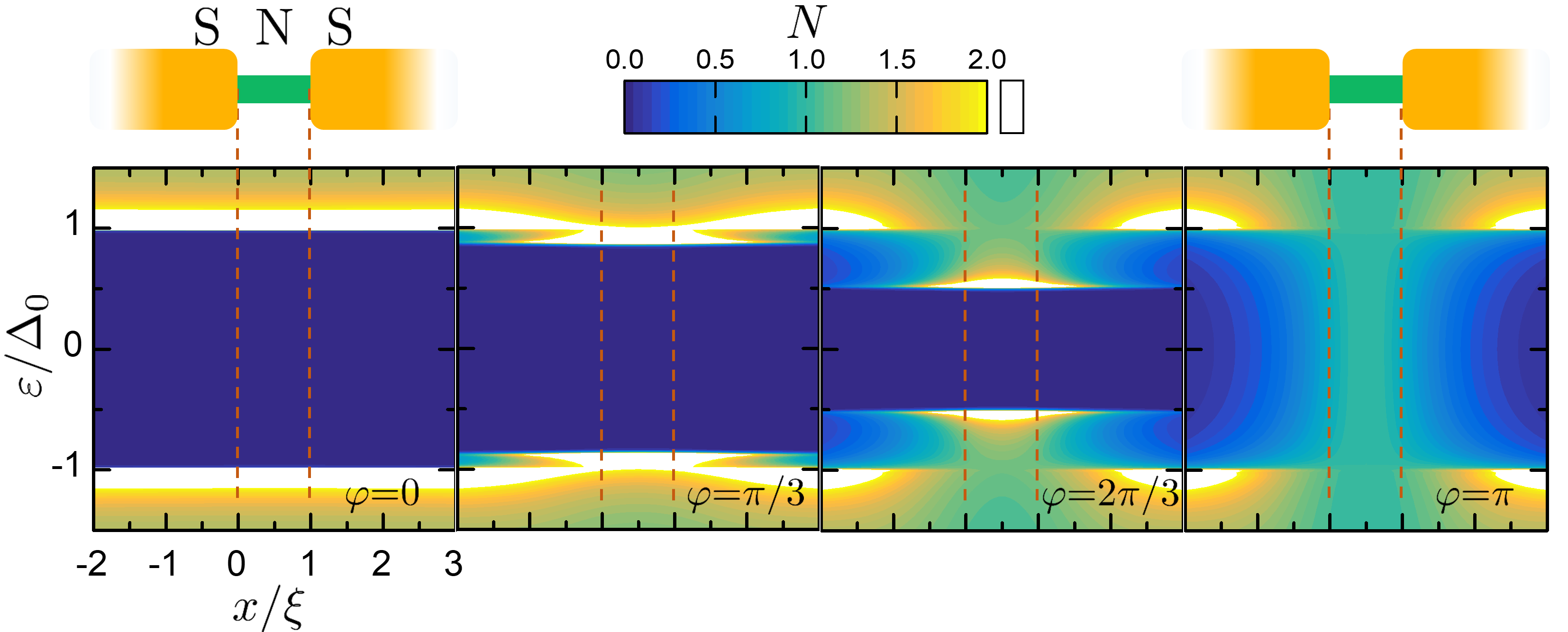}
		\caption{Color plots of the quasi-particle local normalized Density of States (DoS) $N$ in a Superconductor/Normal metal/Superconductor (SNS) junction, versus
			energy $\en$ and position $x$, for~$\ph=0, \pi/3, 2\pi/3,\pi$. The dashed lines separate the S regions (on the sides) to the N region (in the center),
			as shown by the junction sketch. The phase dependence of the DoS is mirrored in a phase-dependence of the junction entropy $S$. The numerical calculation has been obtained within the quasi-classical methods of Reference \cite{virtanen2017SNS} with $a=10,l=0.1,\Delta(T\to 0)=\Delta_0$}
		\label{fig:Proximity}
	\end{figure}

	A numerical example of the phase-dependence of the local normalized DoS $N$ in a junction is reported in Figure \ref{fig:Proximity}, within the quasi-classical
	methods of Reference \cite{virtanen2017SNS}, calculated for a junction with parameters $l=0.1$ and $a=10$, $\Delta(T\to0)=\Delta_0$. The color plots show the evolution of the
	normalized local DoS $N$ versus energy $\en$ and spatial position $x$ for four values of $\ph$ from $0$ to $\pi$. The blue area corresponds to the gapped
	part of the local DoS; the white area is the saturation color that is associated to the divergence of the DoS at the gap edges. The position is normalized
	to the coherence length $\xi$: as shown in the first panel, the central zone $x \in [0,1]$ coincides with the N weak link, while the lateral zones are
	the superconducting leads. At $\ph=0$, the DoS is homogeneous and is approximatively given by the BCS form
	\begin{equation}
	N_{\rm BCS}(\en,T) =  \Re \frac{|\en|}{\sqrt{\en^2 - \Delta^2(T)}} \, \, .
	\label{eq:BCSDoS}
	\end{equation}
	The spatial homogeneity is due to the fact that the calculation involves a short junction, otherwise the induced minigap would have been smaller than $\Delta_0$
	\cite{virtanen2017SNS, heikkila2002, giazotto2011}. Increasing $\ph$, the induced minigap shrinks till the complete closure at $\ph=\pi$. It is possible
	to appreciate also the inverse proximity effect in the S leads, outside the stripe delimited by the red dashed lines in Figure \ref{fig:Proximity}.
	
	The phase dependence of the quasi-particle DoS implies a phase dependence of the junction entropy. The total entropy is \cite{grosso2000,virtanen2017SNS,rabani2008,rabani2009,BCS1957}
	
	\begin{equation}
	S(\ph,T) = \int_{\rm Vol}  {\cal S}(\ph,T,\mathbf{r}) \dd V
	\label{eq:Entropy1}
	\end{equation}
	\begin{equation}
	{\cal S}(\ph,T,\mathbf{r}) = -4 {\cal N}_\mathbf{r} \int _{-\infty} ^{\infty} N({\mathbf{r}},\en,\ph) f(\en,T)\log(f(\en,T))\dd \en \, \, 
	\label{eq:Entropy2}
	\end{equation}
	where ${\cal N}_\mathbf{r}$ is ${\cal N}_S$ or ${\cal N}_N$ whether $\mathbf{r}$ is in the leads or the weak link. 
	
	{At this point, we have two ways to calculate the entropy variation $\delta S(\ph,T)$. One consists in calculating $\delta S(\ph,T)$ from $I(\ph,T)$ exploiting the Maxwell relation through Equations (\ref{eq:MaxwellRelation})--(\ref{eq:ElEnDefinition}). The~other way is by means of Equations (\ref{eq:Entropy1})--(\ref{eq:Entropy2}) given by the statistical argument above concerning the quasi-particle density of states. It is a non-trivial result that the two approaches give results in \mbox{agreement \cite{virtanen2017SNS,kos1999, kosztin1998}}. This is an equilibrium thermodynamic feature due to the fact that the equilibrium supercurrent is carried by the Andreev Bound States (ABS), whose spectral density is non zero below the superconducting gap $|\en|< \Delta(T)$ \cite{hammer2007,heikkila2002}. The quasi-particle DoS and the ABS spectral density are both functions of $\ph$, ensuring that the two approaches are equivalent.}
	
	We conclude this discussion by calculating $S_0$. As discussed in Section \ref{sec:JunctionAsTDSystem}, this quantity can not be obtained by the Maxwell
	relation (\ref{eq:MaxwellRelation}), constituting hence an undetermined function of the temperature $T$ in Equation (\ref{eq:Stot}). However, $S_0$ can
	be determined with a statistical mechanics approach. Given the assumptions of Section \ref{sec:Model} of short junction $l\lesssim 1$ and $a\gg 1$, the local normalized
	DoS at $\ph=0$ is given by the BCS expression (\ref{eq:BCSDoS}) \cite{virtanen2017SNS}. Hence
	\begin{equation}
	S_0(T) = -4 V {\cal N}_S  \int _{-\infty} ^{\infty}  \Re \frac{|\en|}{\sqrt{\en^2 - \Delta^2(T)}} f(\en,T)\log(f(\en,T))\dd \en \, \, .
	\label{eq:BCSEntropy}
	\end{equation}
	In obtaining this expression from (\ref{eq:Entropy2}), we neglected that ${\cal N}_S\neq {\cal N}_N$ in general. However, since the junction volume is
	negligible respect to the total volume, we have approximated the prefactor with ${\cal N}_S V_{\rm leads}+{\cal N}_N V_{\rm weak\,L}\approx V {\cal N}_S $.
	
	Below, we discuss this result within the full dependence of the total entropy $S$ on $\ph$ and $T$.
	
	\subsection{Kulik-Omel'yanchuk Theory}
	The Kulik-Omel'yanchuk theory, whose assumptions have been introduced in Section \ref{sec:Model}, provides the CPR \cite{kulik1975,golubov2004,heikkila2002}
	
	\begin{equation}
	I(\ph,T) = \frac{\pi \Delta(T)}{e R_N} \cos \left(\frac{\ph}{2}\right) \int ^{\Delta(T)} _{|\Delta(T) \cos(\ph/2)|} \frac{1}{\sqrt{\en^2 - \Delta^2(T)\cos^2(\ph/2)}}
	\tanh\left(\frac{\en}{2 T}\right) \dd \en \, \, .
	\label{eq:KOCPR}
	\end{equation}
	This expression \cite{heikkila2002} is equivalent to the Matsubara summation form presented in the first paper about the KO CPR\cite{kulik1975}. Here we
	adopt the integral form that allows to find simple closed expressions in the limit $T\ll \Delta_0$.
	
	In the zero-temperature limit $T\to 0$, the KO CPR reduces to \cite{kulik1975}
	\begin{equation}
	\SupC(\ph,T=0) = \frac{\pi \Delta_0}{e R_N}\cos\left(\frac{\ph}{2}\right) \artanh \left(\sin \frac{\ph}{2}\right) \, \, .
	\label{eq:KOT0}
	\end{equation}
	We use as scale for the supercurrent the critical current $I_c$ at $T=0$, obtained by maximizing (\ref{eq:KOT0}). Numerical maximization returns that $I_c$
	is
	\begin{equation}
	\SupC_{\rm c} = \frac{\kappa \pi \Delta_0}{2 e R_N} \,\, 
	\label{eq:IcT0}
	\end{equation}
	where $\kappa\approx 1.33$. The maximum is placed  at phase $\ph \approx 1.97\approx0.63 \pi$.
	
\begin{figure} [t]
		\centering
		\includegraphics[width=0.95\textwidth]{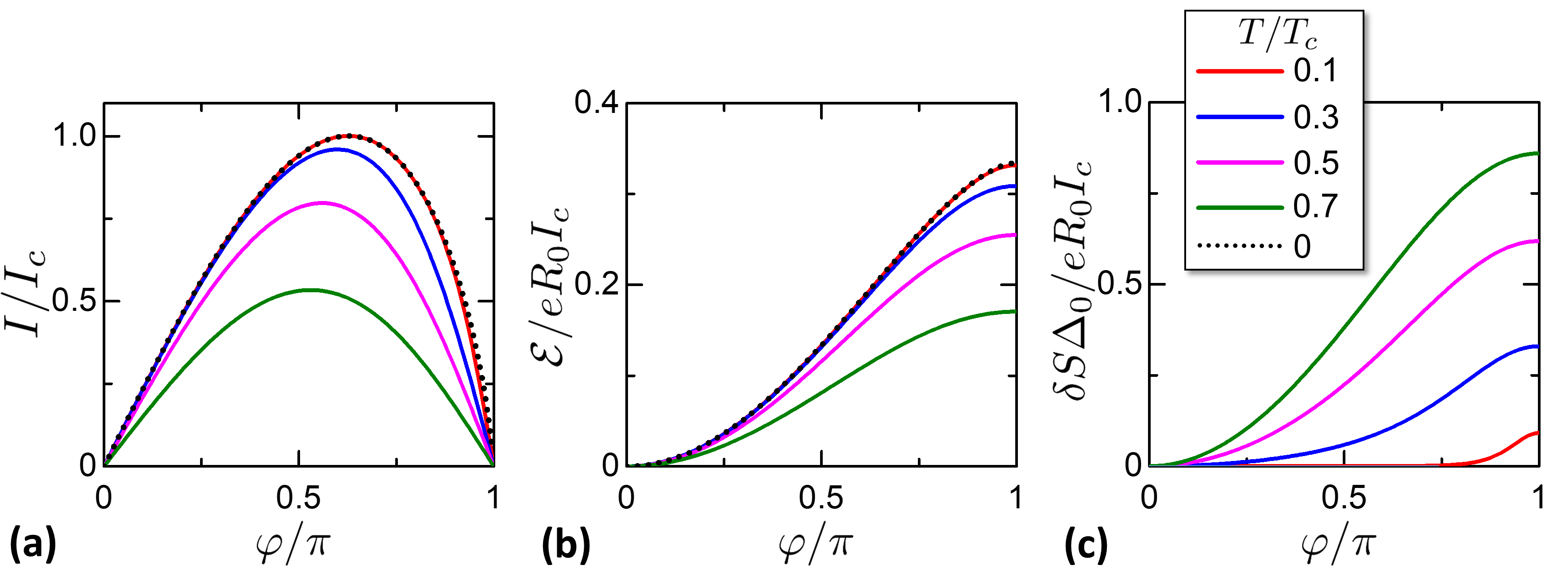}
		\caption{Characteristics of the KO theory, reported versus phase $\ph$ for chosen temperatures $T$ in legend. (\textbf{a}) Supercurrent $\SupC(\ph,T)$,
			in Equation (\ref{eq:KOCPR}). The dotted curve at $T=0$ is given by Equation~(\ref{eq:KOT0}). ({\bf b}) Electric Energy $\ElEn(\ph,T)$, in Equation
			(\ref{eq:KOElEn}). The dotted curve at $T=0$ is given by Equation (\ref{eq:ElEnT0}). ({\bf c})~Entropy variation $\delta S(\ph,T)$, in Equation (\ref{eq:deltaS}).}
		
		\label{fig:KOChars}
	\end{figure}

	The KO CPR is shown in Figure \ref{fig:KOChars}a, normalized to $I_c$. The $T=0$ curve in Equation (\ref{eq:KOT0}) is plotted in black dotted. As one can see the supercurrent
	decreases versus temperature, passing from a skewed shape to a more sinusoidal shape \cite{golubov2004}.
	
	According to the prescription given in (\ref{eq:ElEnDefinition}), the associated Josephson energy to the KO CPR is
	\begin{equation}
	\ElEn(\ph,T) = \frac{R_0}{R_N} \int _{|\Delta(T)\cos(\ph/2)|} ^{\Delta(T)}
	\log \left[
	\frac{\Delta(T) |\sin(\ph/2)|+\sqrt{\en^2-\Delta^2(T)\cos^2(\ph/2)}}{\sqrt{\Delta^2(T)-\en^2}}
	\right]
	\tanh\left( \frac{\en}{2T}\right) \dd \en \, \, 
	\label{eq:KOElEn}
	\end{equation}
	where $R_0 = h/2e^2\approx \SI{12.9}{\kilo\ohm}$ is the inverse of the conductance quantum.
	\begin{comment}
	Sum version of ElEn
	\begin{equation}
	\ElEn(\ph,T) = \frac{\hbar T}{e^2 R_N} \sum_{\omega>0} \arcsin^2
	\left(\frac{\Delta(T)\sin(\ph/2)}{\sqrt{\omega^2 + \Delta(T)^2}}\right) \, \, .
	\label{eq:KOElEn}
	\end{equation}
	\end{comment}
	
	The characteristics of $\ElEn(\ph,T)$ are plotted in Figure \ref{fig:KOChars}b. Being the integral of the supercurrent, the~Josephson energy increases
	versus temperature. At $T=0$, $\ElEn$ reduces to
	\begin{equation}
	\ElEn(\ph,T=0) = \frac{e R_0 \Delta_0}{2 e R_N} \left[
	\log\left(1-\sin^2 \frac{\ph}{2}\right) + 2\artanh\left(\sin\frac{\ph}{2}\right)\sin\frac{\ph}{2}
	\right] \, \, .
	\label{eq:ElEnT0}
	\end{equation}
	The maximum Josephson energy is $\ElEn_0=\ElEn(\ph=\pi,T=0)$, given by
	\begin{equation}
	\ElEn_0 =  \frac{\log 4}{2} \frac{R_0}{R_N} \Delta_0 = \frac{\log 4}{\kappa \pi} e R_0 I_c \, \, 
	\label{eq:ElEn0}
	\end{equation}
	that is about $\ElEn_0 \approx 0.33 eR_0I_c$.
	
	From $\ElEn$ it is possible to calculate $\delta S$. Figure \ref{fig:KOChars}c reports the entropy variation $\delta S(\ph,T)$ calculated numerically with
	$\delta S(\ph,T)=-\partial_T \ElEn(\ph,T)$, for chosen temperatures in legend. It can be noticed that $\delta S$ decreases with the temperature, consistently
	with the third law of thermodynamics.
	
	At low temperatures, where $\partial _T \Delta(T)\to0$, a closed form of $\delta S$ can be obtained by the temperature derivative of Equation (\ref{eq:KOElEn}),
	yielding \cite{virtanen2018,heikkila2002,virtanen2017SNS}
	\begin{equation}
	\delta S(\ph,T) = \frac{R_0}{2R_N} \int ^{\Delta_0} _{\Delta_0 |\cos \frac{\ph}{2}|}  \log \left[ \frac{\Delta_0 |\sin (\ph/2)|+\sqrt{\en^2-\Delta_0^2
			\cos^2 (\ph/2)}}{\sqrt{\Delta_0^2-\en^2}} \right] \frac{\en}{T^2} \sech^2 \left(\frac{\en}{2T}\right) \dd \en
	\, \, .
	\label{eq:LowTdS}
	\end{equation}
	
	The behavior of the entropy can be qualitatively grasped with the minigap mechanism. Let us consider a fixed temperature $T$. Hence, the distribution function
	$f\log f$ in Equation (\ref{eq:Entropy2}) has a certain bandwidth of the order $T$. At low temperature $T\ll\Delta_0$ and $\ph=0$, the DoS gap has width
	$\Delta_0$ and the distribution bandwidth is smaller than the gap. Hence, the lack of available states exponentially suppresses the entropy. When $\ph$
	moves from $\ph=0$ to $\ph=\pi$, the minigap shrinks giving new available states for the distribution $f\log f$, increasing the entropy. At $T\ll\Delta_0$
	and short junction, it is approximately $\tilde{\Delta} = \Delta_0 |\cos (\ph/2)|$ \cite{giazotto2011}, the matching phase between the minigap and the distribution
	bandwidth is $2\arccos (T/\Delta_0)$, at which the entropy increases. This is particularly evident in the curve $T=0.1T_c$ in Figure \ref{fig:KOChars}c,
	where $\delta S$ is negligible except close to $\ph\to \pi$.
	
	\subsection{Total Entropy}
	Given the microscopic and the KO CPR considerations of the last subsections, we can study the total entropy, that is
	\begin{equation}
	S(\ph,T) = S_0 (T) + \delta S(\ph,T)
	\label{eq:EntropyStateEquation}
	\end{equation}
	\textls[-20]{where $S_0$ is given by the BCS entropy in equation (\ref{eq:BCSEntropy}) and $\delta S=-\partial_T \ElEn$  where $\ElEn$ is given by expression (\ref{eq:KOElEn}).}

	We note that the first term scales as $\Delta_0 {\cal N}_S V$, while the second as $e R_0 I_c/\Delta_0$.  For this reason, it is convenient to introduce a parameter $\alpha$
	of the system that sets the ratio between these two quantities:
	\begin{equation}
	\alpha = \frac{e R_0 \SupC_{\rm c}}{{\cal N}_S \Delta_0^2 V } \, \, .
	\label{eq:alpha}
	\end{equation}
	$\alpha$ characterizes the relative influence of the phase-dependent term $\delta S$ over the remaining term $S_0$. The~quantity $\alpha$ can be experimentally determined by heat capacity measurements, as explained in Section~\ref{sec:Isophasics}. Moreover, $\alpha$ controls the temperature of a first-order transition to the normal state when $\ph\neq 0$, discussed in detail in Appendix \ref{sec:Limits}.

	Figure \ref{fig:EntropyScheme} reports the total entropy for $\alpha = 0.6$. Different values $\ph$ in the legend are plotted, showing the increase of $S$
	from $\ph=0$ to $\ph=\pi$. The four curves correspond to the DoS states in the frames of Figure \ref{fig:Proximity}. As expected, the closure of the minigap
	from $\ph=0$ to $\ph=\pi$ implies an increase of entropy. The scale of this increase is set by $\alpha$.
	\begin{figure} [t]
		\centering
		\includegraphics[width=0.95\textwidth]{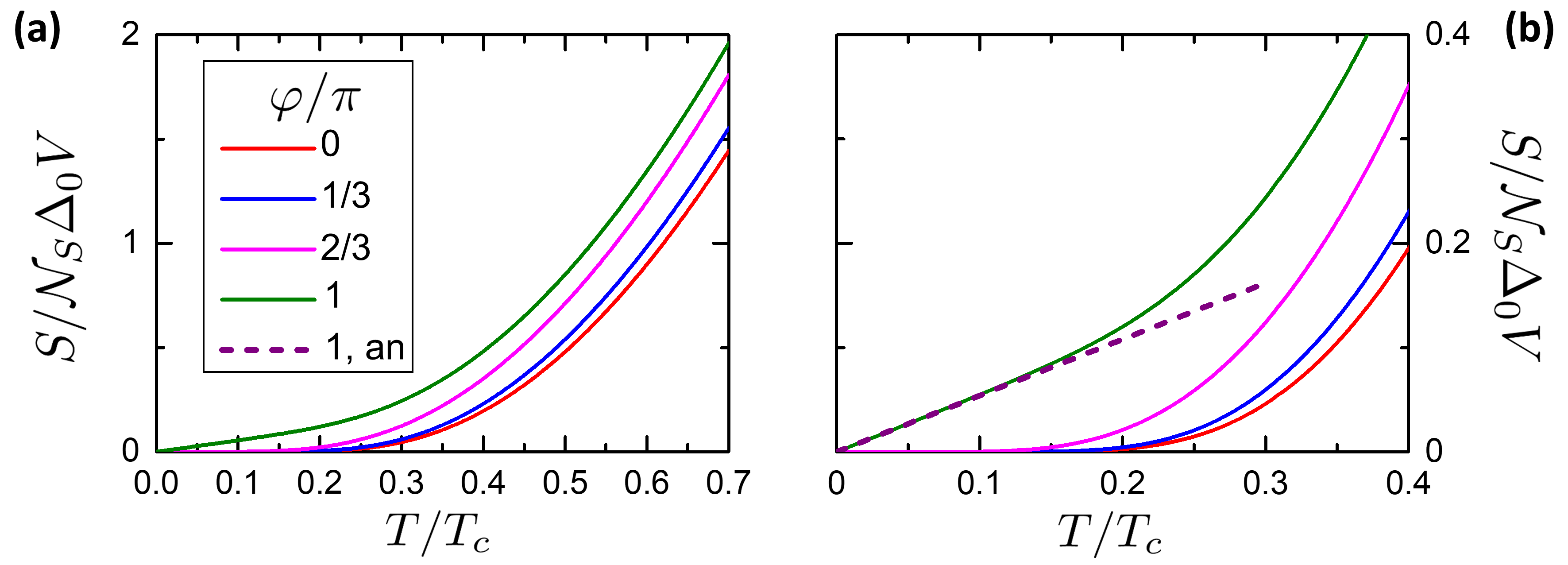}
		\caption{\textls[-10]{Total entropy $S$ of the system for $\alpha=0.6$. ({\bf a}) $S$ versus temperature $T$ for chosen phases $\ph$ in legend. The case $\ph=0$
				correspond to the BCS entropy $S_0(T)$ in Equation (\ref{eq:BCSEntropy}). ({\bf b}) Magnification of panel ({\bf a}) around $T=0.2T_c$, highlighting the
				passage from a exponential suppressed behavior at $\ph=0$ to a linear behavior at $\ph=\pi$. The dashed curve is the analytical low-temperature in expression
				(\ref{eq:deltaSAn}), (\ref{eq:deltaSAn2}).}}
		\label{fig:EntropyScheme}
	\end{figure}
	
	{In the following, the calculations are obtained with $\alpha=0.6$. This value evidences the entropy variation and the related results while keeping a proximized volume negligible respect to the total volume, as shown below in this subsection,  and keeping the unwanted first-order transition above the temperature $0.7T_c$, as discussed in Appendix \ref{sec:Limits}. Considering that ${\cal N}_S\approx\SI{7E46}{\meter^{-3}\joule^{-1}}$ \cite{court2008}, $\alpha=0.6$ corresponds to a ratio $I/V\approx \SI{20}{\milli\ampere\micro\meter^{-3}}$.
	}

	The behavior of the entropy can be studied in more detail at low temperature $T\ll \Delta_0$, where closed expressions can be obtained.
	At $\ph=0$, the DoS has the BCS form in the whole volume of the device, returning the exponentially suppressed behavior of entropy described by the red
	curve in Figure \ref{fig:EntropyScheme}. Hence, at low temperatures $T\ll \Delta_0$ and $\ph=0$ the entropy can be approximated by the expression \cite{degennes1999,
		abrikosov1975}
	\begin{equation}
	S_0(T) \approx \sqrt{2\pi} \sqrt{\frac{\Delta_0}{T}} e^{-\Delta_0/T} V \DoSF \Delta_0 \, \, .
	\label{eq:AnalyticS0}
	\end{equation}
	
	\textls[-10]{At $\ph=\pi$, the minigap is closed and a proximized spatial region around the weak link has a metallic-like DoS. The entropy density (\ref{eq:Entropy2})
		is then exponentially suppressed in the leads and with a linear-in-temperature dependence in the proximized region. This is confirmed by an analytical expression for $\delta
		S$ that can be obtained at low temperatures at $\ph=\pi$. Substituting $\ph=\pi$ in (\ref{eq:LowTdS}) we obtain}
	\begin{equation}
	\delta S(\ph=\pi,T) = -\frac{R_0}{4R_N}\int _0 ^{\Delta_0} \log\left[\frac{\Delta_0-\en}{\Delta_0+\en}\right] \frac{\en}{T^2} \sech\left(\frac{\en}{2 T}\right)\dd\en
	\, \, .
	\end{equation}
	Developing the logarithm around $\en = 0$ as $\log (1 -2\en /(\Delta_0+\en)) \approx -2\en/\Delta_0$ and substituting $\en/T =z$,
	\begin{equation}
	\delta S(\ph=\pi,T) = \frac{R_0}{2R_N} \frac{T}{\Delta_0}\int _0 ^{\Delta_0/T}  z^2 \sech^2 \left(\frac{z}{2}\right)\dd z  \, \, .
	\end{equation}
	For $T\to0$, we obtain
	\begin{equation}
	\delta S (\ph=\pi,T\to 0) = \frac{\pi^2}{3} \frac{R_0}{R_N} \frac{T}{\Delta_0} = \frac{2 \pi}{3} \frac{e R_0 I_c}{\kappa \Delta_0} \frac{T}{\Delta_0} \,
	\, .
	\label{eq:deltaSAn}
	\end{equation}
	The linear behavior of $\delta S (\ph=\pi,T\to 0)$ allows to neglect the exponentially suppressed $S_0$ contribution to the total entropy $S$, allowing the following
	approximation
	\begin{equation}
	S (\ph=\pi,T) \approx \delta S(\ph=\pi,T) \, \, .
	\label{eq:deltaSAn2}
	\end{equation}
	Figure \ref{fig:EntropyScheme}b reports the low-temperature behavior of the entropy for $0\leq T \leq 0.4T_c$. The $\ph=0$ and $\ph=\pi$ curves show  the exponentially
	suppressed  and linear behavior respectively. The purple dashed curve report the analytical expression (\ref{eq:deltaSAn}), revealing a good agreement
	at $T < 0.2T_c$.
	
	A qualitative explanation of the drastic change of the entropy behavior versus  phase difference can be done within the mechanism presented in Section \ref{sec:ProximityMechanism}.
	When $\ph=0$, the local normalized DoS $N(\ph,\en,\mathbf{r})$ is homogeneous over space and gapped according to the BCS expression. As a consequence, the
	entropy density $\cal S$ in (\ref{eq:Entropy2}) is exponentially suppressed and independent on the position $\mathbf{r}$. When $0<\ph<\pi$, the DoS is
	altered: this alteration can be roughly described as an effective proximized volume $\tilde V$, where the DoS is phase-dependent with minigap $\tilde{\Delta}(\ph)$, while in the non-proximized rest of the system the DoS is unchanged with the BCS gap $\Delta_0$. As a consequence, the entropy contribution from the proximized region dominates over the entropy contribution from the non-proximized region, since from the proximized region it is ${\cal S}\propto({\tilde {\Delta}(\ph)/T})^{1/2} e^{-\tilde{\Delta}(\ph)/T}$ while from the non-proximized region it is ${\cal S}\propto({ {\Delta_0}/T})^{1/2} e^{-1/{\Delta_0}}$. This behavior can be noticed in Figure \ref{fig:EntropyScheme}b, comparing the curves at $\ph= \pi/3$ and $\ph= 2\pi/3$
	with the one at $\ph= 0$. The curves at $\ph\neq 0$ show a suppressed region in a low  temperature interval whose width depends on $\tilde \Delta (\ph)$. Finally, when $\ph=\pi$, the induced minigap is closed and the behavior is radically changed from exponentially suppressed to linear.
	
	We conclude this Section with some remarks about $\alpha$. In our treatment, $\alpha$ is a free parameter to be set to get numerical results. However, there
	is a physical upper limit to its value. Maximizing $\alpha$ can be done experimentally by maximizing $I_c$ and minimizing $V$. However, this can not be
	done at will, since our approach is based on the KO theory, that requires as assumption that the two leads are good reservoirs of electron coherence, i.e., 
	the inverse proximity effect by the weak link does not spoil the bulk superconducting properties of the leads. We show the existence of this upper limit
	with the following two arguments.
	
	The first is given by expressing $\alpha$ in terms of the system geometrical properties. Taking into account expression (\ref{eq:IcT0}) for $I_c$, the
	coherence length $\xi^2=\hbar D/\Delta_0$, the S conductivity $\sigma_S= 2 e^2 {\cal N}_S D$ and that the volume is $V\approx A_S L_S$, we have
	\begin{equation}
	\alpha = \kappa\pi^2 \frac{1}{l a} \frac{\xi}{L_S} \, \, .
	\end{equation}
	The requirement $al\gg 1$, given in Section \ref{sec:Model}, implies hence that the only free parameter for increasing $\alpha$ is to decrease the length
	of the ring as most as allowed by the practical geometrical realization.
	
	The second argument about the physical upper limit of $\alpha$ is that the volume $V$ of the system must be in any case bigger than the effective proximized volume $\tilde V$, involving both the weak link and the  in inverse proximized leads. This can be obtained in a qualitatively by considering that when $\ph=0$, the minigap is closed and
	the DoS is modified in a region surrounding the weak link in such a way to return expression (\ref{eq:deltaSAn}), that is linear like a normal metal. The
	volume $\tilde V$ of the proximized region can be estimated by comparing expression (\ref{eq:deltaSAn2}) with the entropy of a normal metal $S_N=2\pi^2{\cal
		N}_S \tilde{V} T/3$:
	\begin{equation}
	\tilde V = \frac{1}{\pi \kappa} \frac{e R_0 I_c}{\Delta_0^2 {\cal N}_S} = \pi \xi A_S \frac{1}{al}\, \, .
	\label{eq:EffectiveProximity}
	\end{equation}
	This expression suggests that the inverse proximized region is present in the leads for a length $\xi /al$.
	
	The volume of the proximized region does not coincide with the weak link region. In particular, they scale differently on the junction length $L_N$, since
	the volume of the weak link is $\propto L_N A_N$, while the proximized volume is $\tilde{V} \propto I_c \propto A_N/L_N$. This point shows that the proximized
	region is not confined in the weak link but is extended also in the leads owing to the inverse proximity effect \cite{virtanen2017SNS}.
	
	By imposing that the proximized volume is smaller than the system volume, $\tilde V \ll V$, Equation~(\ref{eq:EffectiveProximity}) yields
	\begin{equation}
	\alpha \ll \pi \kappa \approx 4.18 \, \, .
	\end{equation}
	
	In our calculations, $\alpha=0.6$, corresponding to a ratio $\tilde V / V \approx 0.15$. Finally, another argument that estimates the physical upper limit of
	$\alpha$ concerns the fact that a high ratio of $I_c/V$ decreases the critical temperature of the system. {This point is discussed in details in Appendix \ref{sec:Limits}.} 
	
	\section{Thermodynamic Processes}
	\label{sec:ThermodynamicProcesses}
	In this section we discuss thermodynamic processes focussing on quasi-static situation, meaning that the device passes through a succession of equilibrium states. This condition can be met if one considers a sufficiently slow speed of the process under inspection. This speed is set by the leading (fastest) thermalization mechanism, that is the electron-electron (e-e) interaction.
	
	Here, we study three different thermodynamic processes. The first is an isothermal one, where the phase is changed while the temperature is kept constant. The second is the isophasic where the temperature is changed while the phase is kept constant. Finally, we consider the isentropic process where the phase is changed while the system exchanges no heat with the universe, thus retaining entropy constant during the process. We will give particular attention to processes with phase variation only between $\ph=0$ and $\ph=\pi$, for two reasons.
	First of all, for these two values of the phase difference, the circulating supercurrent is null and we can neglect any inductive contribution from the ring to the total energy when investigating thermodynamic cycles. Secondly, these particular values of $\ph$ admit simple and closed expressions, allowing for a simple and analytical discussion within KO theory.
	
	For a quasi-static thermodynamic process, the heat flow between the initial and the final state can be written as is
	\begin{equation}
	Q = \int _{\cal P} T \dd S \, \, 
	\end{equation}
	and the work released is
	\begin{equation}
	W = -\frac{e R_0}{2\pi}\int _{\cal P} I \dd \ph \, \, 
	\end{equation}
	where the integrals are meant to be line integrals over the path $\cal P$ in the space of the thermodynamic variables. For quasi-static processes, $\cal P$ lies in the surface of the equilibrium states. We will consider the three different paths corresponding to the isothermal, isophasic and isentropic situation. Hereafter, heat and work integrals are defined according to the following sign convention: the work $W$ is positive when the system releases work to the universe, while the heat $Q$ is positive when the system absorbs heat from the universe. According to this convention, the energy conservation over a closed loop path reads $Q-W=0$.
	
	\subsection{Isothermal Process}
	Let us consider a isothermal process from an initial state $i$ at $(\ph_i,T)$ to a final state $f$ at $(\ph_f,T)$. This~can be realized by keeping open the heat valve toward the reservoir sketched in Figure \ref{fig:Sketch}a. For~notation simplicity, here we indicate both the system temperature and the reservoir temperature as $T$, implying that at thermal  equilibrium $T=\overline{T}$, where $\overline{T}$ is the reservoir temperature.
	
	In this case the work released by the system is
	\begin{equation}
	W_{if} = -\frac{e R_0 }{2\pi}\int _{\ph_i} ^{\ph_f} I(\ph,T)\dd \ph =  \ElEn(\ph_i,T) - \ElEn(\ph_f,T) \, \, 
	\end{equation}
	where $\ElEn$ represents the Josephson energy in the junction defined in Equation (\ref{eq:deltaS}). For a process $\ph_i=0\to\ph_f=\pi$, where the universe has to perform a work on the system, the sign of $W_{if}$  is negative, consistently with the convention of $W$ defined above.
	
	The heat absorbed during this process is
	\begin{equation}
	Q_{if} = T(S(\ph_f,T)-S(\ph_i,T)) = T(\delta S(\ph_f,T)-\delta S(\ph_i,T)) \, \, .
	\label{eq:IsothermalHeat}
	\end{equation}
	Heat is absorbed when $\ph$ goes from $0$ to $\pi$, owing to the closure of the minigap. It is worth to note that in the isothermal process we do not explicitly rely on the BCS contribution $S_0(T)$. {Hence, the~thermodynamic consistency requires that an isothermal process must exchange heat. Interestingly,  the~supercurrent is not directly involved in this heat exchange, since the Cooper pair system carries no entropy and the supercurrent flow is dissipationless. Instead, the heat is absorbed by the quasi-particle (excited states of system) from an external system, i.e.,  in our scheme from the external reservoir at fixed temperature $\overline{T}$. If heat is not supplied, the system undergoes an adiabatic transformation, treated in the next subsection. Below in Section \ref{sec:Experimental} we discuss some strategies to measure this heat exchange. }
	
	At low temperature $T\ll \Delta_0$, the heat absorbed and the work released in an isothermal process from $\ph=0$ to $\ph=\pi$ can be calculated exploiting the expression (\ref{eq:deltaSAn}). 
	\begin{equation}
	Q_{if} = T \delta S(\ph=\pi,T) = \frac{2\pi}{3 \kappa} \left(\frac{T}{\Delta_0}\right)^2  e R_0 I_c \, \, 
	\label{eq:IsothermalHeatAn}
	\end{equation}
	where the second equivalence is due to the fact that the temperature is constant during an isothermal~process.
	
	The released work at low temperature is obtained by calculating the expression of $\ElEn$ at low temperature. From Equation (\ref{eq:deltaSAn}), since $\delta S=-\partial_T \ElEn$, it is
	\begin{equation}
	\ElEn (\ph=\pi, T \ll \Delta_0) = \ElEn_0 -\frac{\pi}{3\kappa}\left(\frac{T}{\Delta_0}\right)^2 e R_0 I_c = \left[\frac{\log 4}{\kappa \pi} -\frac{\pi}{3\kappa}\left(\frac{T}{\Delta_0}\right)^2 \right] e R_0 I_c\, \,  
	\label{eq:ReleasedWorkLowT}
	\end{equation}
	where we have used the expression (\ref{eq:ElEn0}) for $\ElEn_0$. The work at low temperature for a $\ph=0\to\pi$ isothermal is
	\begin{equation}
	W_{if} = -\left[\frac{\log 4}{\kappa \pi} -\frac{\pi}{3\kappa}\left(\frac{T}{\Delta_0}\right)^2 \right] e R_0 I_c \, \, .
	\end{equation}
	As expected, the work released scales as the critical supercurrent $I_c$ and increases in module by lowering the temperature. 
	
	\subsection{Isophasic Process and Heat Capacity}
	\label{sec:Isophasics}
	In an isophasic process, the phase difference $\ph$ is kept constant while the temperature is changed. Considering Figure \ref{fig:Sketch}, this can be done by opening the thermal valve toward the reservoir while the threading flux $\Phi$ is fixed. The system passes from its initial temperature $T_i$ to the final temperature $T_f=\overline{T}$.
	
	The work exchanged is then null, since $dW = -e R_0 I \dd \ph/2\pi$. The system exchanges energy only in the form of heat. In a process from $(\ph,T_i)$ to $(\ph,T_f)$, the exchanged heat can be written as
	\begin{equation}
	Q_{if} = \int _{i\to f} T \dd S = T_f S(\ph,T_f)  - T_i S(\ph,T_i) - \int _{T_i} ^{T_f} S(\ph,T) \dd T
	\end{equation}

	At low temperature $T_i,T_f \ll \Delta_0$, using Equations (\ref{eq:AnalyticS0}) and (\ref{eq:deltaSAn}), it is possible to obtain the isophasic heat in a closed form for $\ph=0$ and $\ph=\pi$. At $\ph=0$ we get
	\begin{equation}
	Q_{if} \approx \sqrt{2\pi} \left[ \sqrt{\frac{T_f}{\Delta_0}}e^{-\Delta_0/T_f} - \sqrt{\frac{T_i}{\Delta_0}}e^{-\Delta_0/T_i} \right] V \DoSF \Delta_0
	\label{eq:IsophasicHeatAnPhi0}
	\end{equation}
	while for $\ph=\pi$ we have
	\begin{equation}
	Q_{if} \approx \frac{\pi}{3\kappa} e R_0 \SupC_{\rm c}\left[ \left(\frac{T_f}{\Delta_0}\right)^2 - \left(\frac{T_i}{\Delta_0}\right)^2\right] \, \, .
	\label{eq:IsophasicHeatAn}
	\end{equation}
	
	The heat exchanged in an isophasic process brings naturally to the concept of heat capacity. Indeed the heat exchanged can be expressed as a function of the initial and final temperatures as
	\begin{equation}
	Q_{if} = \int _{T_i} ^{T_f} C(\ph,T) \dd T
	\end{equation}
	where $C(\ph,T)$ is the isophasic heat capacity:
	\begin{equation}
	C (\ph,T) = \left(\frac{\partial Q}{\partial T}\right)_\ph = T \frac{\partial S(\ph,T)}{\partial T} \, \, .
	\label{eq:HeatCapDef}
	\end{equation}
	The importance of the heat capacity relies also on the fact that is an experimental observable quantity. Indeed, by definition, can be measured as the temperature response of the system to a heat pulse. 
	
	From Equations (\ref{eq:IsophasicHeatAnPhi0}) and (\ref{eq:IsophasicHeatAn}) it is evident that the amount of heat exchanged for an isophasic process from $T_i$ to $T_f$ depends on the phase $\ph$ and, hence, the heat capacity is dependent on $\ph$. From these expressions, it is possible to obtain the isophasic heat capacity at low temperature $T \ll \Delta_0$ in a closed form. At $\ph=0$
	\begin{equation}
	C (\ph=0,T) \approx \sqrt{2\pi} \left(\frac{\Delta_0}{T}\right)^{3/2} e^{-\Delta_0/T} V \DoSF \Delta_0 \, \, 
	\label{eq:AnCphi0}
	\end{equation}
	while at $\ph=\pi$
	\begin{equation}
	C (\ph=\pi,T) \approx \frac{2 \pi}{3 \kappa} \frac{e R_0 \SupC_{\rm c}}{\Delta_0} \frac{T}{\Delta_0} \, \, .
	\label{eq:AnC}
	\end{equation}
	Similarly to the entropy, the heat capacity assumes two different behaviour,passing from a suppressed superconducting-like to a linear metallic-like behavior whether the phase is $\ph=0$ or $\ph=\pi$, respectively. Moreover, the ratio of (\ref{eq:AnC}) over (\ref{eq:AnCphi0}) scales like $\alpha$. Hence, an experimental measurement of $C$ at $\ph=0$ and $\ph=\pi$ can be used to get an estimated value for the dimensionless parameter $\alpha$ discussed before.
	
	Here,  a sort of parallelism between the variables $(I,\ph)$ of our system and $(p,V)$ of an ideal gas can be noticed. In the same analogy, the work differential $e R_0 I \dd \ph/2\pi $ plays the role of the $p dV$ differential for a classic gas.
	However, the phase difference $\ph$ variable is $2\pi$ periodic, differently from $V$.

\begin{figure} [t]
		\centering
		\includegraphics[width=0.95\textwidth]{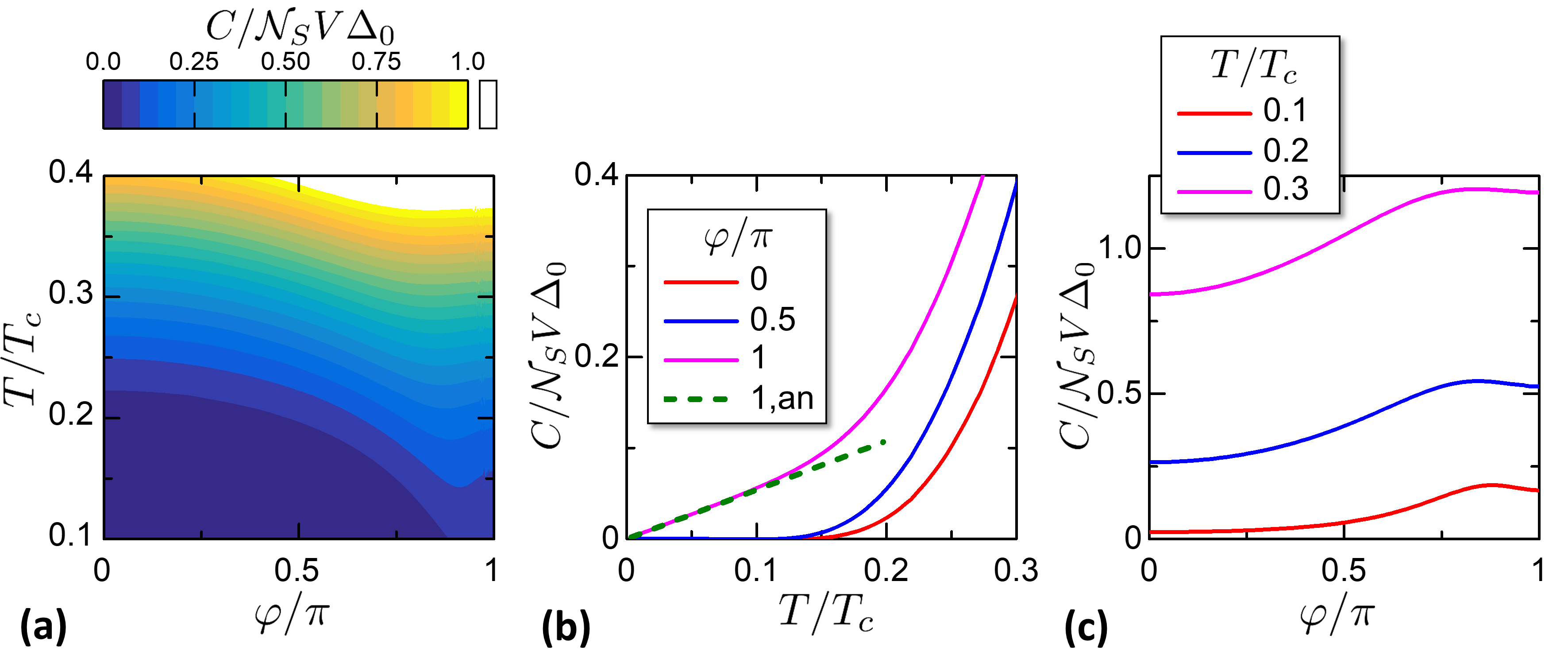}	
		\caption{Isophasic heat capacity properties. ({\bf a}) Map of the isophasic heat capacity $C(\ph,T)$. ({\bf b}) Cuts from panel (\textbf{a}) for chosen phases in legend. The dashed line shows the low temperature expression (\ref{eq:AnC}). ({\bf c}) Cuts from panel (\textbf{a}) for chosen temperatures in legend.}
		\label{fig:HeatCapacity}
	\end{figure}

	In a generic situation, the heat capacity can be evaluated numerically from Equation (\ref{eq:HeatCapDef}). Figure~\ref{fig:HeatCapacity} reports $C(\ph,T)$ for $\alpha=0.6$. Figure 	\ref{fig:HeatCapacity}a is a color plot of $C$ versus $\ph$ and $T$; panels (b) and (c) are cuts of the color plot versus $T$ and $\ph$. Looking at panels (a) and (b) one can note that $C(\ph,T)$ goes from a gapped-like behavior at $\ph=0$ to a linear behavior at $\ph=\pi$. This is confirmed also by the dashed green line plotting the analyitical expression (\ref{eq:AnC}). Since $C$ is the temperature derivative of $S$, its~behavior can be explained qualitatively within the phase-dependent minigap mechanism, in the same fashion provided for the entropy in Section \ref{sec:Thermodynamics}.
	
	\subsection{Isentropic Process}
	\textls[-10]{In a isentropic process the entropy of the system is conserved. Since we are considering quasi-static processes, an isentropic is also adiabatic in the thermodynamic meaning that no heat is exchanged with the universe. Indeed, for a quasi-static process, it holds $dQ=TdS$. The isentropic process can be physically realized when the system is thermally isolated, i.e.,  when the heat valve in Figure \ref{fig:Sketch}a is closed.}  
	
	In a isentropic process, the constraint of constant $S$ implies an implicit relation between the phase and the temperature. Let us consider an isentropic process that starts at the initial state $(\ph_i,T_i)$. During the whole process, $\ph$ and $T$ are related by the implicit equation
	\begin{equation}
	S(\ph_i,T_i) = S(\ph,T) \, \, .
	\label{eq:IsentropicEq}
	\end{equation}
	For an isolated system, the  phase and entropy $(\ph,S)$ are the independent variables that will specify the system state. The temperature is then a function $T(\ph,S)$. In detail, $T(\ph,S)$ decreases by increasing $\ph$ in the interval $0<\ph<\pi$ for fixed $S$. Indeed, since $S(\ph,T)$ is an increasing function in $0<\ph<\pi$ for fixed $T$ and, as a consequence, the temperature of the system must decrease in order to keep $S$ constant. 
	
	In particular, we focus on the isentropic temperature decrease for processes that start at phase $\ph=0$ and $T_i$, where the initial state sets the entropy $S(\ph=0,T_i)$ of the process. For these processes, we~define the temperature decrease $T_f$ implicitly defined in Equation (\ref{eq:IsentropicEq}) as
	\begin{equation}
	T_f(\ph,T_i) = T(\ph,S(\ph_i=0,T_i)) \, \, .
	\end{equation}

\begin{figure} [t]
		\centering
		\includegraphics[width=0.95\textwidth]{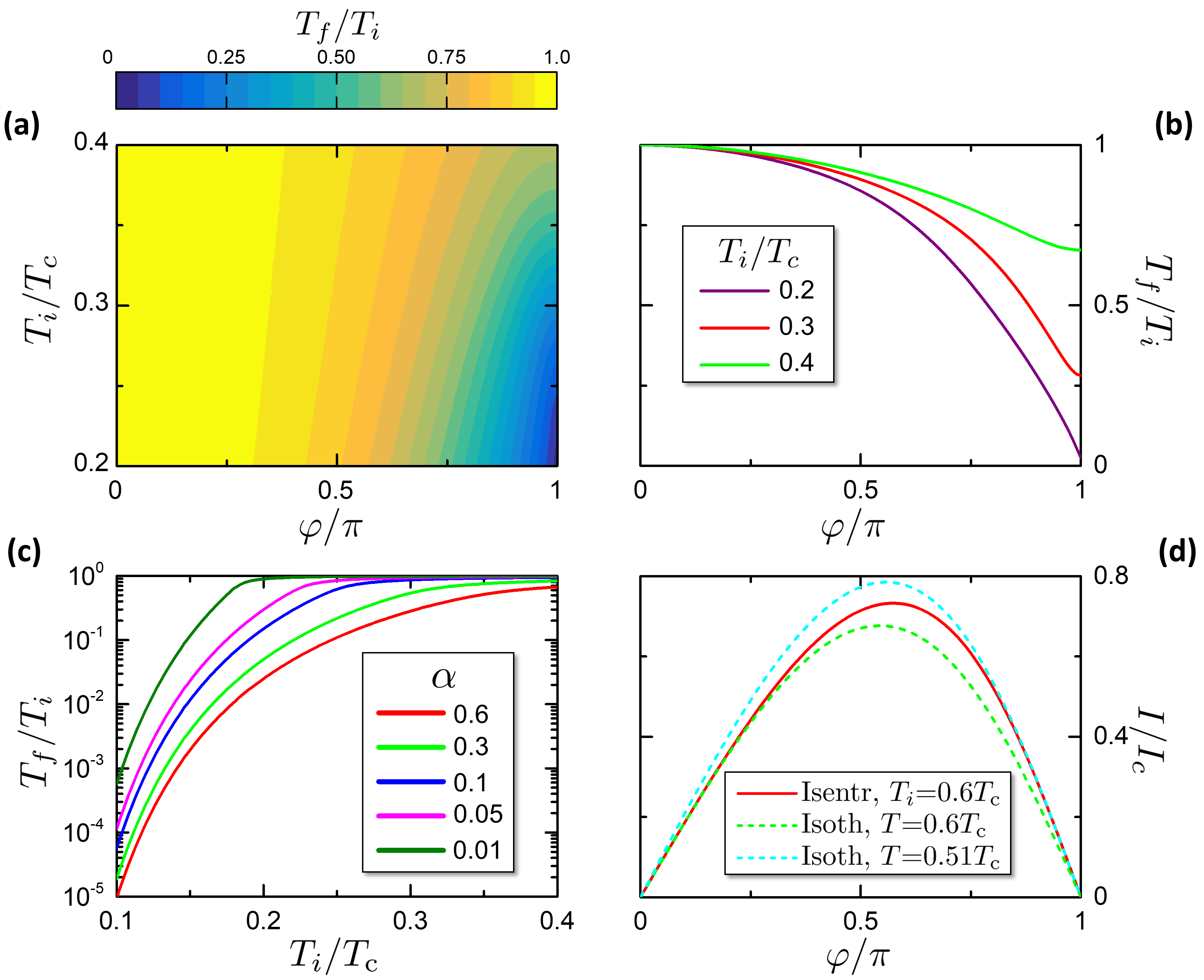}
		\caption{Isentropic processes properties. ({\bf a}) Colormap of the temperature decrease $T_f(T_i,\ph)/T_i$) for an isentropic process from initial temperature $T_i$ at $\ph=0$ to $\ph$. ({\bf b}) Cuts from panel (a) for the chosen temperatures in legend. ({\bf c}) Temperature decrease $T_f/T_i$ for an isentropic process from $(\ph=0,T_i)$ to $(\ph=\pi,T_f)$ for different values of $\alpha$. ({\bf d}) Isentropic current phase relation (red solid curve) across the state $(\ph=0,T_i=0.6T_c)$, for $\alpha=0.6$. For comparison, the dashed curves report two isothermal current phase relations at $T = 0.6T_c$ and $T = 0.51T_c$ (see legend).}
		\label{fig:IsentropicProcess}
	\end{figure}

	In Figure \ref{fig:IsentropicProcess}a,b we report the quantity $T_f(\ph,T_i)/T_i$, i.e.,  the relative temperature decrease, for a system with $\alpha = 0.6$. $T_f/T_i$ is enhanced toward low $T_i$, since for $T \ll \Delta_0$ the behaviors of $S(\ph=0,T)$ and $S(\ph=\pi,T)$ are strongly different: the former is exponentially suppressed while the latter is linear (see Figure \ref{fig:EntropyScheme}a,b).
	
	A closed expression for $T_f/T_i$ can be obtained for $T\ll \Delta_0$ by exploiting equations (\ref{eq:AnalyticS0}),(\ref{eq:deltaSAn}):
	\begin{equation}
	\frac{T_f}{T_i} = \frac{3\kappa}{\alpha \sqrt{2\pi}} \left(\frac{\Delta_0}{T_i}\right)^{3/2}e^{-\Delta_0/T_i}\, \, .
	\label{eq:TDecreaseAn}
	\end{equation}
	
	The isentropic cooling is reminiscent of the adiabatic cooling process typical for the expansion of an ideal gas. The analogy goes forward when discussing in terms of available states. Indeed, when the gap reduces to closure in the process $\ph=0\to\ph=\pi$ the number of available states increases so that the temperature decreases to keep the entropy constant. The same thing happens in the case of an adiabatic expansion of a gas, where the position states are increased by the volume increase.
	
	Figure \ref{fig:IsentropicProcess}c plots $T_f/T_i$ versus $T_i$ for different values of $\alpha$. The relative cooling is more effective for higher $\alpha$, since higher values of $\alpha$ correspond to a stronger weight of the proximized region, where the gap can be tuned, over the phase independent superconducting leads. In particular, $T_f/T_i \propto \alpha^{-1}$ as shown in Equation (\ref{eq:TDecreaseAn}). However, the passage from a gapped to a gapless state yields a strong temperature cooling even for moderate values of $\alpha$, provided that $T_i$ is low enough.
	
	It is interesting to investigate how the CPR of a junction is modified by the assumption that during the change of the phase difference $\ph$ the entropy remains constant. For sake of simplicity, here we focus on isentropic CPR where the entropy is set by the initial state by $S(\ph_i=0,T_i)$. In such case the Josephson current can be calculated substituting the temperature $T_f(\ph,T_i)$ in the isothermal CPR (\ref{eq:KOCPR}):
	\begin{equation}
	I_S(\ph,T_i) = I(\ph,T_f(\ph,T_i)) \, \, .
	\label{eq:ISIT}
	\end{equation}
	In  Figure \ref{fig:IsentropicProcess}d we show the comparison between the isothermal and isentropic CPR for $\alpha=0.6$ for the initial temperature $T_i=0.6T_c$. It is worthy to notice that the isothermal CPR depends only on the nature of the junction, while for the isentropic case there is also a dependence on $\alpha$, which includes the total volume of the system. The dashed lines are isothermal curves at the initial temperature $T_i=0.6T_c$ and the final temperature $T_f(T_i,\ph=\pi)=0.51T_c$. We can notice that the two isothermal respectively overlap the isentropic at $\ph \to 0$ and $\ph\to \pi$. Moreover, the fact that the isentropic curve lies between the two isothermal curves indicates that the temperature is between the initial and final temperatures, since $T$ evolves from $T_i$ to $T_f$ during the isentropic process.
	
	The work for an isentropic $\ph=0\to \pi$ with initial temperature $T_i$ is given by 
	\begin{equation}
	W_{if} = -\frac{eR_0}{2\pi} \int _0 ^{\ph} I_S(\ph',T_i) \dd \ph' =-\frac{eR_0}{2\pi} \int _0 ^{\ph} I(\ph',T_f(\ph',T_i)) \dd \ph' \, \, .
	\end{equation}
	Since the isentropic CPR is constrained between the isothermal CPRs at $T_i$ and $T_f$, i.e.,  $I(\ph,T_i)<I_S(\ph,T_i)<I(\ph,T_f)$ as shown in Figure \ref{fig:IsentropicProcess}d, the isentropic work is equally constrained between the isothermal works at $T_i$ and $T_f$.
	
	\section{Thermodynamic Cycles}
	\label{sec:ThermodynamicsCycles}
	\textls[-10]{The combination of different thermodynamic processes, studied in the previous section, allows constructing thermodynamic cycles.
		In this section, we present two possible examples of thermodynamic cycles that {can be built based on} the various processes discussed above.
		In particular, we focus on two cycles that we call Josephson-Otto cycle and Josephson- Stirling cycle, thanks to their analogy with classic thermodynamic counterpart. We first explain their implementation and then we discuss their performances.}
	
\begin{figure} [t]
		\centering
		\includegraphics[width=0.95\textwidth]{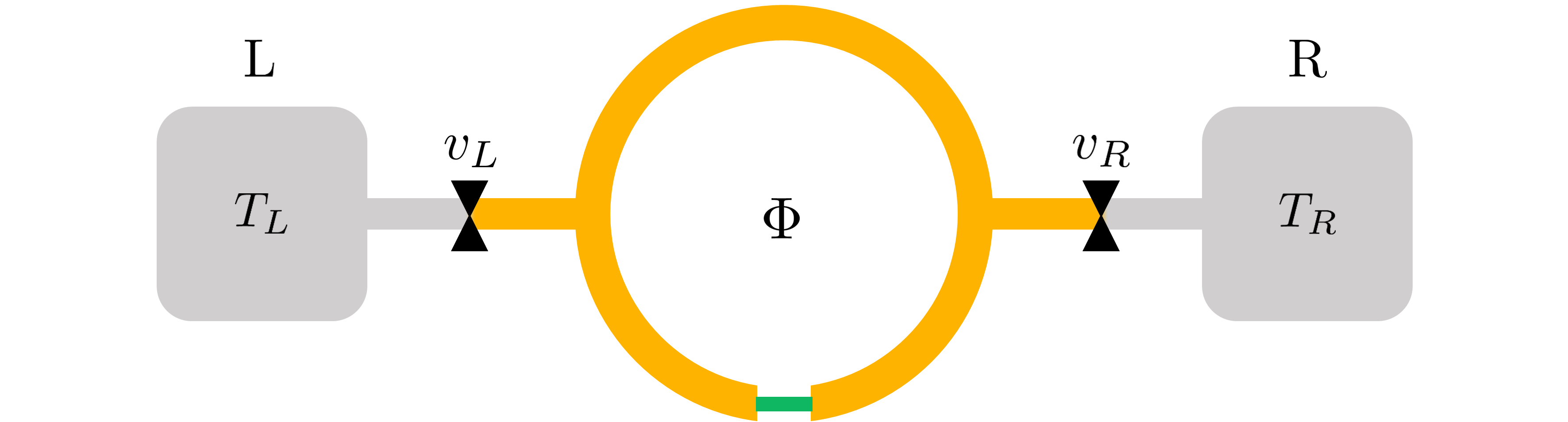}
		\caption{Sketch of the system connected to two reservoirs, identified as Left Reservoir (L) and Right Reservoir (R), through two heat valves $v_L,v_R$ respectively. Thermodynamic cycles can be implemented varying configurations between different temperatures $T_L$ and $T_R$, achieving also opposite operational modes such as engine or refrigerator configurations (see text).
		}
		\label{fig:Sketch2}
	\end{figure}

	To this aim, we consider the hybrid system attached to two different reservoirs, identified as Left Reservoir (L) and right Reservoir (R), in the sketch of Figure \ref{fig:Sketch2}. The two reservoirs are at  fixed temperature $T_j$ and can release heat $Q_j$ to the system through a heat channel controlled by a heat valve $v_j$, where the subscript $j$ can be $L$ or $R$, respectively.
	We consider $Q_j$ positive when the heat flows from the reservoir to the system, in agreement with the sign convention defined in Section \ref{sec:ThermodynamicProcesses}. Thereafter, we~will study cycle characteristics as a function of the temperatures $(T_L,T_R)$. In particular, we will show that there are regions of $(T_L,T_R)$ where the cycles can operate as engine or refrigerator. The reservoir roles depend on the operating mode: when a cycle operates as engine, the two reservoirs play the role of the Hot Reservoir (HR) and Cold Reservoirs (CR), at temperatures $T_{hr}>T_{cr}$ respectively. In a cycle, the system absorbs an amount $Q_{hr}$ from the HR and releases  $|Q_{cr}|<Q_{hr}$ heat to the CR. In practical systems, the cold reservoir can be constituted by the ambient, i.e.,  the large substrate thermalized to the cryostat, while the hot reservoir can be a heated subsystem, like a large metallic pad heated by Joule effect.
	
	Conversely, when the cycle is considered as a refrigerator, the two reservoirs play the role of the Cooled Subsystem (CS) and Heat Sink (HS), at temperatures $T_{cs}<T_{hs}$ respectively. In a cycle, the~system absorbs an amount of $Q_{cs}$ from the CS and releases $|Q_{hs}|> Q_{cs}$ to the HS. In practice, the~CS is an isolated subsystem from which the heat is extracted, where the heat capacity is assumed to be large enough to consider the CS as a reservoir within one cycle. The CS can be constituted of a metallic pad that can be used as cooled substrate for nanodevices. In practical systems, the heat sink is typically constituted by the ambient, i.e.,  the substrate thermalized to the cryostat in our device.
	
	The cycle performances are characterized by inspecting several figures of merit. In the case of the engine we investigate the work released per cycle $W$ and its efficiency, defined as
	\begin{equation}
	\eta = \frac{W}{Q_{hr}} \, \, .
	\end{equation}
	{This quantity is physically limited by the Carnot efficiency}
	\begin{equation}
	\eta_C = 1-\frac{T_{cr}}{T_{hr}} \, \, .
	\end{equation}
	In the following subsections, we show $W$ and $\eta$ versus both the temperatures $T_{cr},T_{hr}$. We discuss in detail the dependence of $W$ and $\eta$ as a function of $T_{hr}$ for fixed $T_{cr}$, since in real systems it is most likely possible to tune the HR temperature while the CR temperature $T_{cr}$ is fixed by the ambient.
	
	In the refrigerator mode, the figures of merit we consider are the extracted heat $Q_{cs}$ from the CS per cycle, and the Coefficient of Performance (COP), defined as
	\begin{equation}
	\COP = \frac{Q_{cs}}{|W|} \, \, .
	\end{equation}
	Like the efficiency, the COP is limited physically by the Carnot COP limit
	\begin{equation}
	\COP _C = \frac{T_{cs}}{T_{hs}-T_{cs}} \, \, .
	\end{equation}
	In the following subsections, we show $Q_{cs}$ and the COP versus both temperatures $T_{cs},T_{hs}$. We discuss in detail the dependence of $Q_{cs}$ and the COP as a function of $T_{cs}$ for fixed $T_{hs}$, since in real systems the HS temperature $T_{hs}$ is given by the ambient and can not be tuned, while $T_{cs}$ decreases from the ambient temperature in the refrigeration process.
	
	Notice that the work $W$ and the heat extracted $Q_{cs}$ are quantities defined per cycle. Hence, at~cycling frequency $\nu$, the engine returns a Power $\dot W = W \nu$ and the refrigerator returns a Cooling Power ${\rm CP} = Q_{cs} \nu$.
	
	\subsection{Josephson-Otto Cycle}
	
\begin{figure} [t]
		\centering
		\includegraphics[width=0.95\textwidth]{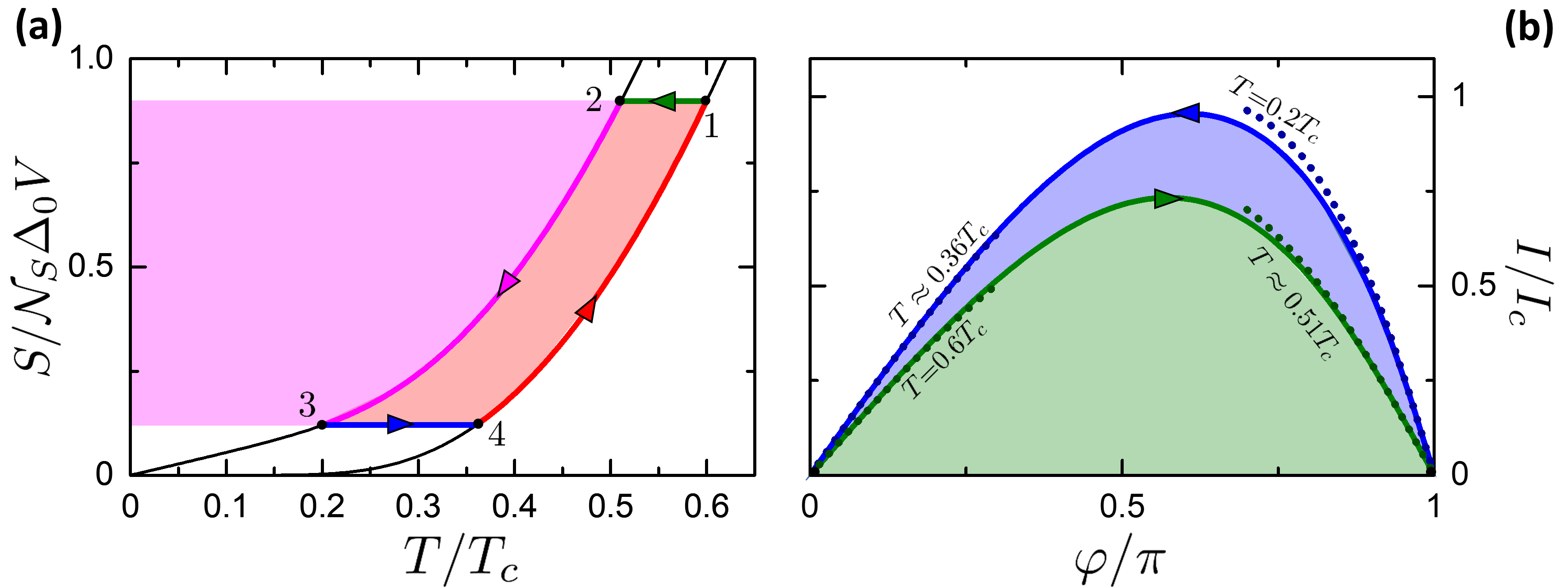}
		\caption{\textls[-10]{Otto cycle scheme. The example considers an engine from a hot reservoir at $0.6T_c$, cold reservoir at $0.2T_c$ and $\alpha=0.6$. ({\bf a}) Scheme in the $(T,S)$ plane. The colored areas help for the discussion in the text of the heat exchanges. ({\bf b}) Scheme in $(\ph,\SupC)$ plane. Of the four processes of the Otto cycle, only the two isentropic are visible, since the two isophasics are collapsed at the points $(\ph=0,I=0)$ and $(\ph=\pi,I=0)$. The colored areas help for the discussion in the text of the work exchanges. For completeness, the dotted curves represent partial isothermal CPRs at the labelled temperature in the plot.}}
		\label{fig:OttoScheme}
	\end{figure}

	Here we study the Josephson-Otto cycle, by starting with the engine mode for sake of simplicity. The Josephson-Otto engine is described by the scheme in Figure \ref{fig:OttoScheme}, where the panels a and b show respectively the processes in the $(T,S)$ and $(\ph,I)$ planes. The cycle is constituted by two isentropic processes, i.e.,  $\bf 1\to2$ and $\bf 3\to4$, and two isophasic processes, i.e.,  $\bf 2\to3$ and $\bf 4\to1$, see Figure \ref{fig:OttoScheme}. We choose by convention that the state $\bf 1$ and $\bf 3$ are  thermalized to the R and L reservoir, respectively. In this way, the R and L reservoirs play  respectively the role of the HR and CR.
	
	The cycle is given by the succession of the following processes:
	\begin{itemize} [leftmargin=*,labelsep=5.5mm]
		\item {\bf Isentropic $\bf 1 \to 2$}. All thermal valves are closed to make the system thermally isolated. The~system is driven from the state  $(\ph_1=0, T_1=T_R)$ to $(\ph_2=\pi, T_2)$, where $T_2 = T_f(\ph=\pi,T_1)$. In~this process the universe spends a work $|W_{12}|$ ($W_{12}<0$ according to the convention defined in Section \ref{sec:ThermodynamicProcesses}). $|W_{12}|$ is  represented by the green area in Figure \ref{fig:OttoScheme}b. No heat is exchanged, $Q_{12}=0$.
		\item {\bf Isophasic $\bf 2 \to 3$}. By opening the thermal valve \LV, the system goes from the state $(\ph=\pi,T_2)$ to $(\ph=\pi,T_3=T_L)$. The system releases heat $|Q_{23}|$ to the left reservoir (magenta area in Figure~\ref{fig:OttoScheme}a). No work is performed, $W_{23}=0$.
		\item {\bf Isentropic $\bf 3 \to 4$}. All thermal valves are again closed to make the system thermally isolated. The system is driven from the state  $(\ph_3=\pi, T_3=T_L)$ to $(\ph_4=0, T_4)$. By construction, if~$T_2>T_L$ then it is $T_4< T_R$. In this process the system returns a work $W_{34}$ ($W_{34}>0$ according to our convention), represented by the sum of the green and blue areas in Figure \ref{fig:OttoScheme}b. No heat is exchanged, $Q_{34}=0$.
		\item {\bf Isophasic $\bf 4 \to 1$}. By opening the thermal valve \RV, the system goes from the state $(\ph=0,T_4)$ to $(\ph=0,T_1=T_R)$. The system absorbs heat $Q_{41}$ from the reservoir at $T_R$ (magenta+pink area in Figure \ref{fig:OttoScheme}a). No work is performed, $W_{41}=0$.
	\end{itemize}
	
\begin{figure} [t]
		\centering
		\includegraphics[width=0.95\textwidth]{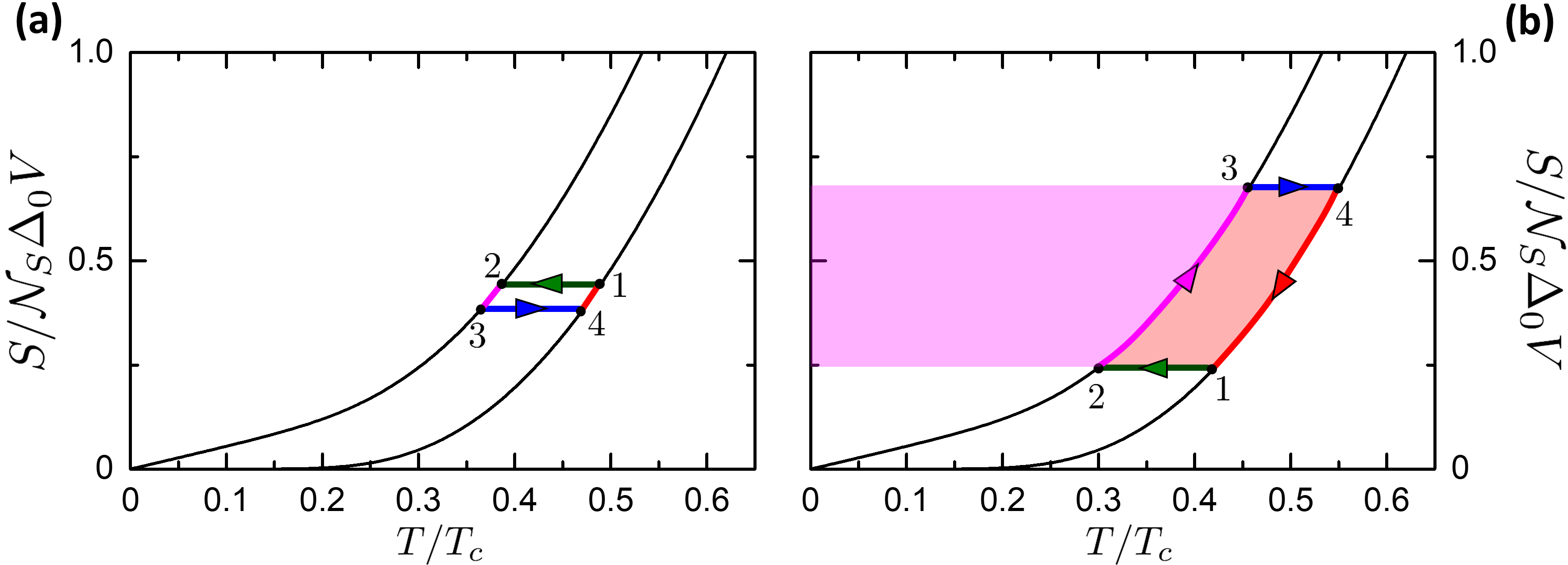}
		\caption{Particular cases of the Otto cycle on $T_L,T_R$. ({\bf a}) Approaching the degenerate case of $T_f(T_R) = T_L$. ({\bf b}) Otto cycle as refrigerator for $T_f(T_R) < T_L$.}
		\label{fig:OttoDegenerate}
	\end{figure}

	The total work released per cycle is
	\begin{equation}
	W =   W_{12} + W_{34} \, \, 
	\label{eq:OttoWorkPerCycle}
	\end{equation}
	The heat $Q_{hr}$ absorbed from the HR (correspondent to R) is
	\begin{equation}
	Q_{hr} = Q_R = Q_{41} \, \, .
	\end{equation}
	From the two schemes presented in  Figure \ref{fig:OttoScheme} it can be noticed that the cycle operates as an engine if $T_2 > T_3$. This condition requires that $T_L < T_f (\ph=\pi,T_R)$, i.e.,  a temperature gap between the two reservoirs is required. When $T_L$ approaches $T_f(\ph=\pi, T_R)$ the cycle tends to the degenerate case reported in Figure \ref{fig:OttoDegenerate}a, where the two adiabatic curves tend to superimpose. Also in the $(\ph,I)$ plane the two adiabatic curves tend to superimpose, meaning that the net work is $W=0$ at $T_L =T_f(\ph=\pi, T_R)$. On the contrary, if $T_L > T_f(\ph=\pi, T_R)$, the cycle is inverted as in Figure \ref{fig:OttoDegenerate}b. In this case, the cycle works as a refrigerator and the work is $W<0$, i.e.,  made by the universe on the system.

	Hence, the curve in the plane $(T_L,T_R)$ where $W=0$ can be defined as the characteristic curve of the Otto cycle. It separates the regions where the cycle is in the engine or refrigerator mode and it is given by the equation
	\begin{equation}
	T_L = T_f (\ph=\pi,T_R) \, \, .
	\label{eq:OttoCharCurve}
	\end{equation}
	
	Close to the characteristic curve, in the case shown in Figure \ref{fig:OttoDegenerate}a, it is evident that $Q_{L}$, $Q_{R}$ tend to zero but their ratio tends to $Q_R/Q_L\to T_R/T_L$. This property is exploited below to calculate the limits of $\eta$ and COP close to the characteristic curve.
	
	Let us consider the refrigerator mode in the case $T_L > T_f (T_R)$, represented in Figure \ref{fig:OttoDegenerate}b. In this case, the cycle is clockwise and operates as a refrigerator. The two reservoirs play a different role: the R reservoir represents the Heat Sink, while the L one represents the Cooled Subsystem. The~case $T_L > T_f (T_R)$  coincides with the following cycle
	\begin{itemize}[leftmargin=*,labelsep=5.5mm]
		\item {\bf Isentropic $\bf 1 \to 2$}. All thermal valves are closed to make the system thermally isolated. The system is driven from the state at the ambient temperature $(\ph_1=0, T_1=T_R)$ to $(\ph_2=\pi, T_2)$, where $T_2 = T_f(\ph=\pi,T_1)$ . In this process, the universe spends a work $|W_{12}|$ ($W_{12}<0$ for of Section \ref{sec:ThermodynamicProcesses}). No heat is exchanged, $Q_{12}=0$.
		\item {\bf Isophasic $\bf 2 \to 3$}. By opening the thermal valve \LV, the system goes from the state $(\ph=\pi,T_2)$ to $(\ph=\pi,T_3=T_L)$, removing the heat $Q_{23}$ from the CS (magenta area in Figure \ref{fig:OttoDegenerate}b). No work is performed, $W_{23}=0$.
		\item {\bf Isentropic $\bf 3 \to 4$}. All thermal valves are closed. The system is driven from the state  $(\ph_3=\pi, T_3=T_L)$ to $(\ph_4=0, T_4)$. Now, $T_4>T_R$. In this process, the system returns a work $W_{34}$. No~heat is exchanged, $Q_{34}=0$.
		\item {\bf Isophasic $\bf 4 \to 1$}. By opening the thermal valve \RV, the system goes from the state $(\ph=0,T_4)$ to $(\ph=0,T_1=T_R)$. The system releases heat $Q_{41}$ to the reservoir at $T_R$, since $T_4>T_R$, which correspond to the magenta+pink area in Figure \ref{fig:OttoDegenerate}b. The temperature $T_4$ plays an analogous role of the hot heat exchanger that is present in the refrigerators. No work is performed, $W_{41}=0$.
	\end{itemize}
	In the refrigerator mode, the work released is still given by $W=W_{12}+W_{34}$. The heat $Q_{cs}$ absorbed by the CS is 
	\begin{equation}
	Q_{cs} = Q_L = Q_{23} \, \, .
	\end{equation}

\begin{figure}[!t]
		\centering
		\includegraphics[width=0.9\textwidth]{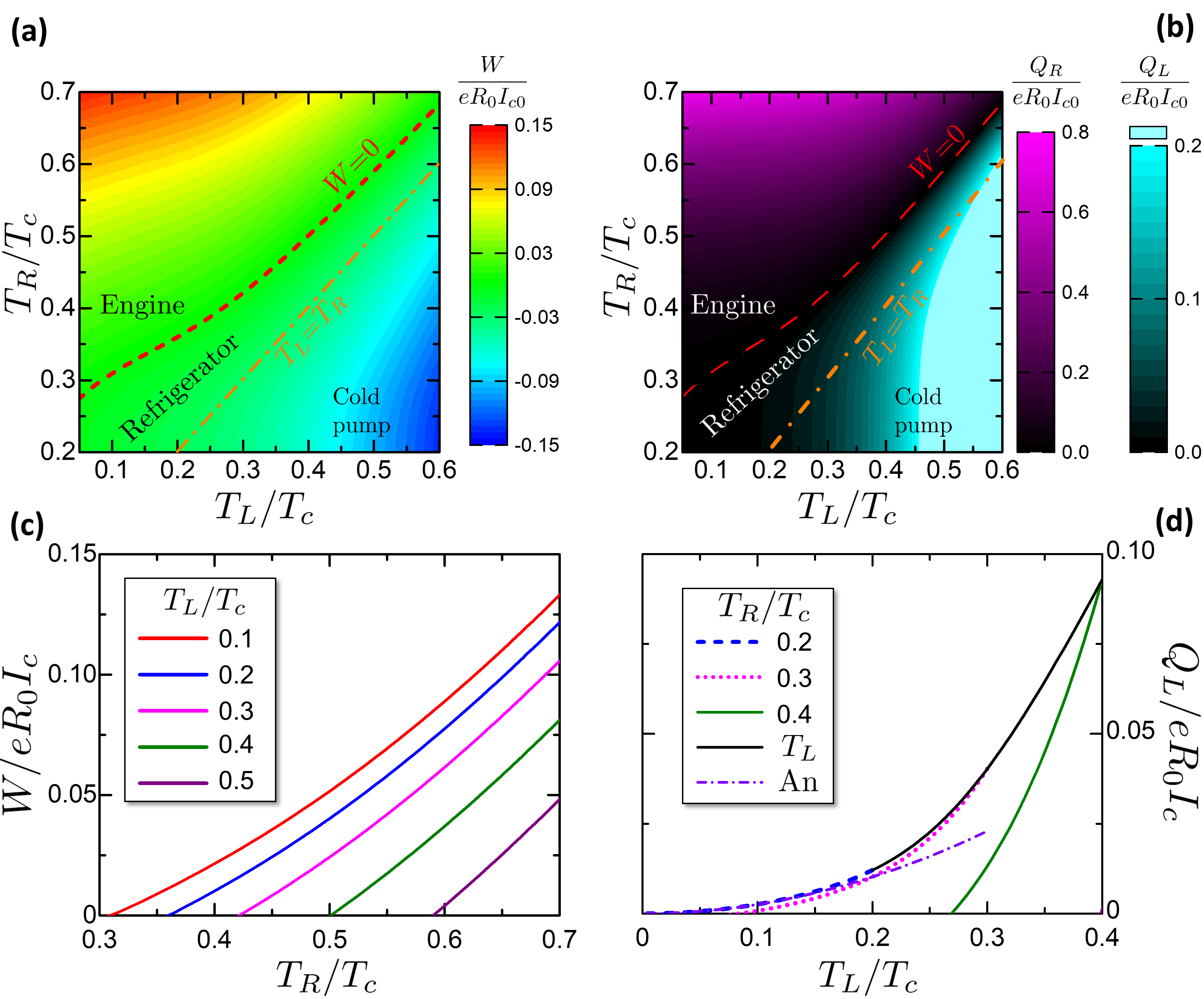}
		\caption{
			({\bf a}) Work released in a Josephson-Otto cycle as a function of $(T_L,T_R)$. The dashed red curve, given by Equation (\ref{eq:OttoCharCurve}), reports $W=0$ and separates the region where the cycle operates as engine or refrigerator. ({\bf b}) Heat absorbed in a Josephson-Otto cycle. As an engine, the heat $Q_R$ from the Hot reservoir is represented by the R reservoir. As a refrigerator, the heat $Q_L$ from the CS is represented by the L reservoir. The dash-dotted line represents the thermal equilibrium $T_L=T_R$, below which the system is a cold pump. ({\bf c}) Cuts of the work in panel (\textbf{a}) versus the Hot Reservoir temperature $T_R$ for fixed temperatures $T_L$ of the Cold Reservoir. ({\bf d}) Cuts of the absorbed heat  versus the CS temperature $T_L$ for fixed temperatures $T_R$ of the Heat Sink. The black solid curve reports the absorbed heat at $T_L=T_R$. The violet dash-dotted curve reports the analytical result of Equation (\ref{eq:OttoQcsAn}).  The curves have been obtained with $\alpha=0.6$.
		}
		\label{fig:OttoWQ}
	\end{figure}

	Figure \ref{fig:OttoWQ} is a summary of the work released $W$ and the heat absorbed $Q_{hr}$ and $Q_{cs}$. Panels a,b are color plots of these quantities versus $(T_L, T_R)$. The dashed red curve represents the characteristic curve defined in Equation (\ref{eq:OttoCharCurve}), corresponding to $W=0$. Above it, for $T_L< T_f(\ph=\pi,T_R)$, the cycle operates as engine, while below it ($T_L> T_f(\ph=\pi,T_R)$) the cycle works as refrigerator. The orange dot-dashed curve reports the thermal equilibrium $T_L=T_R$. We can notice that below this curve, i.e.,  for $T_R<T_L$, there is a region where work is spent by the universe to pump heat from the L reservoir (the hotter one) to the R reservoir (the colder one). Hence, work is spent to perform a process that can be performed spontaneously. We define this region as a cold pump, following the definition given in References \cite{dickerson2016, bizarro2017}. 
	
	Figure \ref{fig:OttoWQ}c reports the released work versus the HR temperature $T_R$ for different values of the CR temperature $T_L$ as reported in the legend. The curves reach the value zero corresponding to the characteristic curve plotted in Figure \ref{fig:OttoWQ}a. We observe that the general trend of the work is to increase with the temperature difference $T_R-T_L$ between the two reservoirs. The order of magnitude of the work per cycle is $\sim 0.1 e R_0 I_c$. 
	
	Figure \ref{fig:OttoWQ}d reports the absorbed heat $Q_{cs}=Q_L$ versus the CS temperature $T_L$ for different values of the HS temperature $T_R$. The black curve reports the case of the heat absorbed $Q_{cs}$ at $T_L=T_R$. The curves with fixed $T_R$ are limited on the right at $T_L=T_R$, to not include the Cold Pump case, see color plots in Figure \ref{fig:OttoWQ}. The curves with fixed $T_R$ goes to zero in correspondence of the characteristic curve, defining the minimum achievable temperature of the refrigerator. The refrigerator can not physically cool below the minimum achievable temperature, since the absorbed heat reaches $Q_{cs}=0$.
	
	The black curve reporting $Q_{cs}$ at $T_L=T_R$ is important since it reports the heat absorbed per cycle when the refrigerator starts to operate at the thermal equilibrium. Hence, for a cycling frequency $\nu$, the~corresponding cooling power $\dot Q_{cs} = Q_{cs}\nu$ for $T_L=T_R$ gives the maximum heating power leakage that the refrigerator can sustain. If the heat leakage is above the cooling power at the thermal equilibrium, no net refrigeration can be accomplished. The heat absorbed per cycle has the same order of the work per cycle, $\sim 0.1 e R_0 I_c$. 
	
	It is possible to find an analytic expression for $Q_{cs}$ valid for $T_R,T_L \ll \Delta_0$. Considering the scheme in Figure \ref{fig:OttoDegenerate}b, it can be noticed that at low temperature the heat absorbed by the CS is ruled by the purple area defined by the linear expression of entropy in Equations (\ref{eq:deltaSAn}) and (\ref{eq:deltaSAn2}). Approximating the $T_2$ temperature to 0, due to the strong isentropic cooling at low temperatures, the $Q_{cs}$ is given at the leading order by the CS temperature 
	\begin{equation}
	Q_{cs} \approx \frac{\pi}{3\kappa} \left(\frac{T_L}{\Delta_0}\right)^2 e R_0 I_c \, \, .
	\label{eq:OttoQcsAn}
	\end{equation}
	This expression is plotted in Figure \ref{fig:OttoWQ}d as a violet dash-dotted curve. The agreement with the numerical results is good at $T_L<0.2T_c$, corresponding to the agreement range in Figure \ref{fig:EntropyScheme}b.

\begin{figure}[t]
		\centering
		\includegraphics[width=0.95\textwidth]{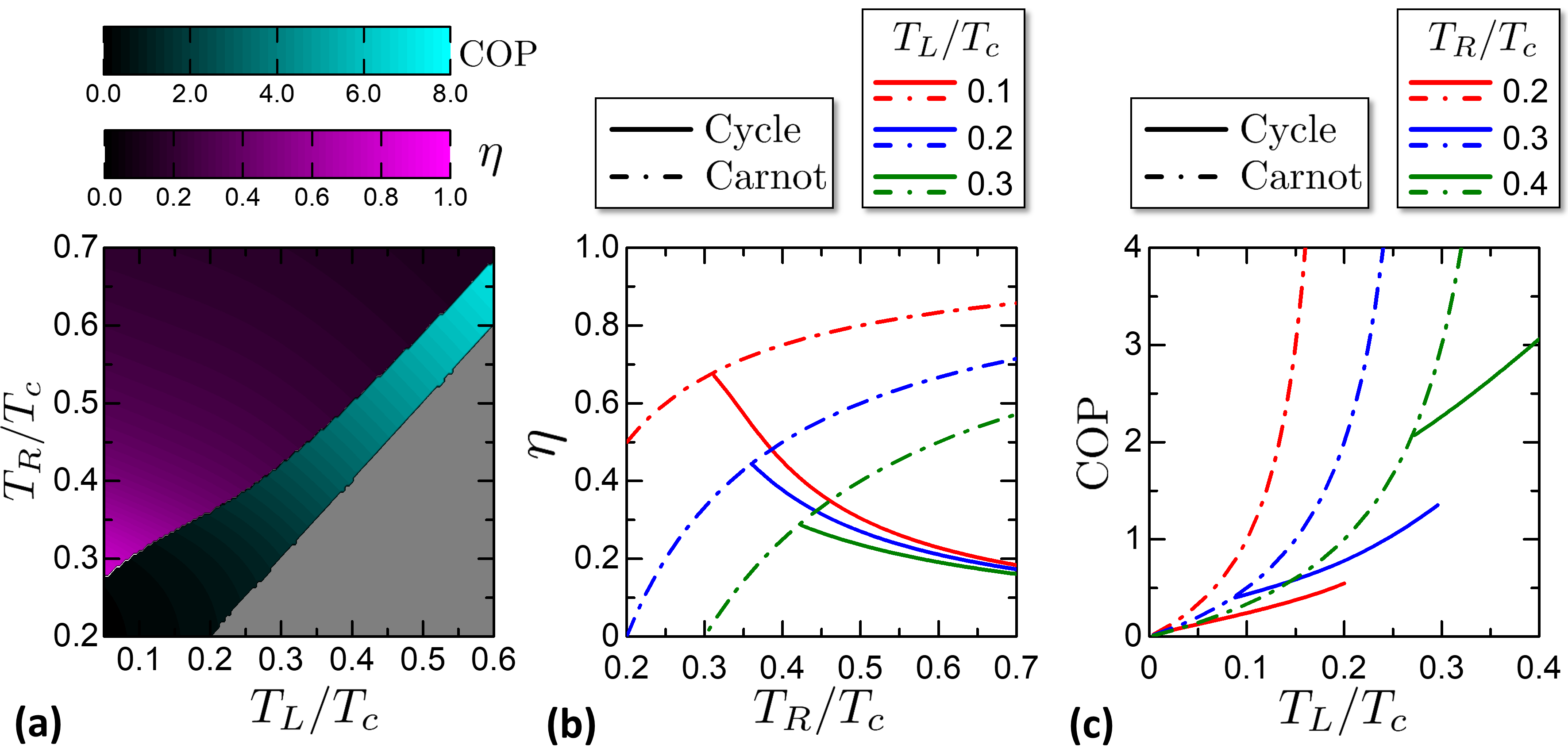}
		\caption{Efficiency and COP of the Otto machine. ({\bf a}) Color plot of $\eta$ and $\COP$ versus $(T_L,T_R)$, with~different color palettes. The gray region represents the state where the cooled subsystem temperature is above the heat sink temperature. ({\bf b}) Cuts of Otto cycle efficiency $\eta$ versus $T_R$ for chosen $T_L$ in legend. The dot-dashed line reports the Carnot limit to efficiency. The curves end at the Otto characteristic curve, Equation (\ref{eq:OttoCharCurve}), where the efficiency reaches the Carnot limit. ({\bf c}) Cuts of Otto cycle COP versus $T_L$ for chosen $T_R$ in legend. The dot-dashed line report the Carnot limit to COP. The curves are limited on the right by the thermal equilibrium state $T_L=T_R$; on the right, the curves are limited by the Otto cycle characteristic curve. On this curve, the COP reaches the COP Carnot limit.}
		\label{fig:OttoEfficiency}
	\end{figure}

	From the characteristics of $W$, $Q_{hr}$, $Q_{cs}$ in Figure \ref{fig:OttoWQ} it is possible to calculate numerically the engine efficiency and the refrigerator COP.  Figure \ref{fig:OttoEfficiency} reports the efficiency and the COP for the studied Otto engine. Figure \ref{fig:OttoEfficiency}a shows a color plot of $\eta$, COP versus $T_L,T_R$. The two quantities are confined respectively in the engine and refrigerator regions of $(T_L,T_R)$. The gray area corresponds to the Cold Pump case. Figure \ref{fig:OttoEfficiency}b reports cuts of the efficiency $\eta$ versus the Hot Reservoir temperature $T_R$ for chosen ambient temperatures $T_L$. The curves end on the left in correspondence of the Otto characteristic curve, where the efficiency saturates at the Carnot limit. Figure \ref{fig:OttoEfficiency}c reports cuts of the COP versus the CS temperature $T_L$ for chosen HS temperatures $T_R$, showing the evolution of the COP when the CS is cooled  down toward the minimum achievable temperature,  that delimits the COP curves on the left. The COP curves are limited on the right by the thermal equilibrium state $T_L=T_R$, where the COP reaches the theoretical Carnot limit. 
	An interesting property of the Josephson-Otto cycle is that close to the characteristic curve, both $\eta$ and the COP reach the Carnot limit, even though the work released or the heat absorbed goes to zero. This point can be explained by referring to the degenerate case of Figure \ref{fig:OttoDegenerate}a. Close to the characteristic curve, the quantities $Q_{L},Q_{R}$ tend to zero but their ratio tends to $|Q_L/Q_R| \to T_L/T_R$. Exploiting the energy conservation $Q_L+Q_R-W=0$, it is 
	\begin{equation}
	\eta (T_L\to T_f(\ph=\pi,T_R)) = 1+ \frac{Q_L}{Q_R} \to 1 -\frac{T_L}{T_R}\, \, 
	\end{equation}
	that is the Carnot limit. With similar considerations, we obtain the analogous limit for the COP:
	\begin{equation}
	\COP (T_L\to T_f(\ph=\pi,T_R)) =  \frac{Q_L}{Q_R+Q_L} \to \frac{T_L}{T_L-T_R} \, \, .
	\label{eq:COP}
	\end{equation}
	
	\subsection{Josephson-Stirling Cycle}
	In this section, we analyze another possible thermodynamic cycle, i.e.,  a Josephson-Stirling cycle, that has in practice several practical applications, in particular as refrigerator \cite{mungan2017}. The Stirling cycle is a different combination of the studied processes, being built with two isochorics and two isophasics. In~an ideal gas system, it consists of two isochoric heat addition/rejection processes and two isothermal (compression + expansion). Real Stirling engines are eventually equipped by regenerators that increase the efficiency \cite{deacon1994, wheatley1986}; here we study the simple case without the regenerators.

\begin{figure}[t]
		\centering
		\includegraphics[width=0.95\textwidth]{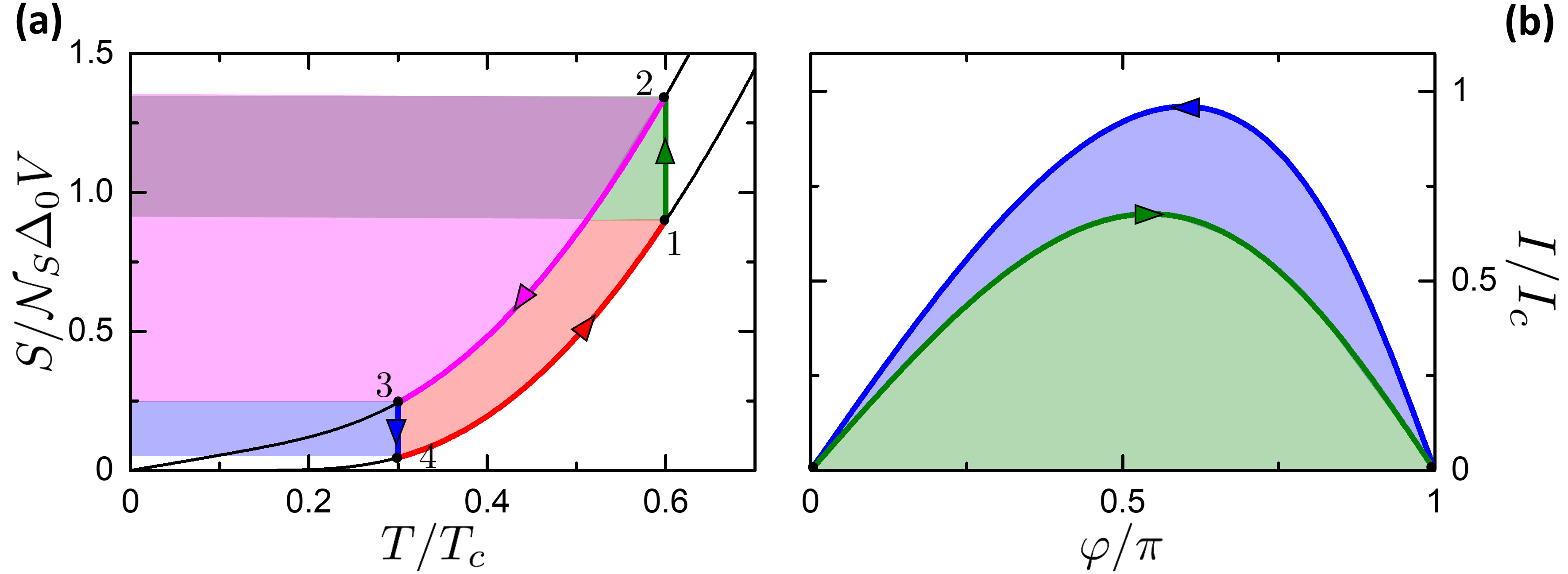}
		\caption{Josephson-Stirling cycle scheme. The plotted example concerns an engine between a hot reservoir $T_R=0.6T_c$ and a cold reservoir $T_L = 0.3T_c$ and $\alpha=0.6$.  ({\bf a}) Scheme in the $(T,S)$ plane. The~colored areas help for the discussion in the text about the exchanged heats. (\textbf{b}) Scheme in $(\ph,{\cal I})$ plane. Of the four processes of the Josephson-Stirling cycle, only the two isothermals are visible, since the two isophasics are collapsed at the points $(\ph=0,I=0)$ and $(\ph=0,I=0)$. The colored areas help for the discussion in the text about the exchanged works.
		}
		\label{fig:StirlingScheme}
	\end{figure}

	\textls[-10]{First of all, let us consider the engine case and then move to the refrigerator one. The~Josephson-Stirling engine is described by the scheme in Figure \ref{fig:StirlingScheme}, where panels a and b show respectively the processes in the $ST$ diagram and $I\ph$ diagram. The cycle is constituted by two isothermal processes ($\bf 1\to2$ and $\bf 3\to4$ in Figure \ref{fig:StirlingScheme}) and two isophasic processes ($\bf 2\to3$ and $\bf 4\to1$ in Figure \ref{fig:StirlingScheme}). The states $\bf 1,2$ and $\bf 3,4$ are respectively thermalized to the right and left reservoirs. When operating as Stirling engine, the left and right reservoirs play the roles of ambient and heat source respectively.}
	
	In summary, the Josephson-Stirling engine is given by the succession of the following processes:
	\begin{itemize}[leftmargin=*,labelsep=5.5mm]
		\item {\bf Isothermal $\bf 1 \to 2$}. \textls[-10]{The thermal valves \RV is open and \LV is closed, so that the system is in thermal contact with the right reservoir. The system is driven from the state  $(\ph_1=0, T_1=T_R)$ to $(\ph_2=\pi, T_2=T_R)$. Here a work is spent $|W_{12}|$ represented by the green area in Figure \ref{fig:StirlingScheme}b. The heat $Q_{12}$ is absorbed from the reservoir, represented by the green + dark purple area in Figure \ref{fig:StirlingScheme}a.}
		\item {\bf Isophasic $\bf 2 \to 3$}. By closing \RV and opening \LV, the system goes from the state $(\ph=\pi,T_2)$ to $(\ph=\pi,T_3=T_L)$. The system releases heat $Q_{23}$ to the left reservoir, represented by the light purple + dark purple area. No work is performed, $W_{23}=0$.
		\item {\bf Isothermal $\bf 3 \to 4$}. The valves are kept in the same state: \RV open and \LV closed. The system is driven from the state  $(\ph_3=\pi, T_3=T_L)$ to $(\ph_4=0, T_4=T_L)$. In this process the system returns a work $W_{34}$ represented by the sum of the green and blue areas in Figure \ref{fig:StirlingScheme}b. The heat $|Q_{34}|$ is released to the left reservoir, represented by the blue area in Figure \ref{fig:StirlingScheme}a.
		\item {\bf Isophasic $\bf 4 \to 1$}. By closing \LV and opening \RV, the system goes from the state $(\ph=0,T_4)$ to $(\ph=0,T_1=T_R)$. The system absorbs the heat $Q_{41}$ from the reservoir at $T_R$, given by the sum of the areas in blue, red and light purple in Figure \ref{fig:StirlingScheme}a. No work is performed, $W_{41}=0$.
	\end{itemize}
	The total work per cycle is given by $W = W_{12}+W_{34}$. The heat absorbed from the Hot R (represented by the R reservoir) is
	\begin{equation}
	Q_{hr} = Q_R = Q_{12}+Q_{41} \, \, .
	\end{equation}

\begin{figure}[t]
		\centering
		\includegraphics[width=0.95\textwidth]{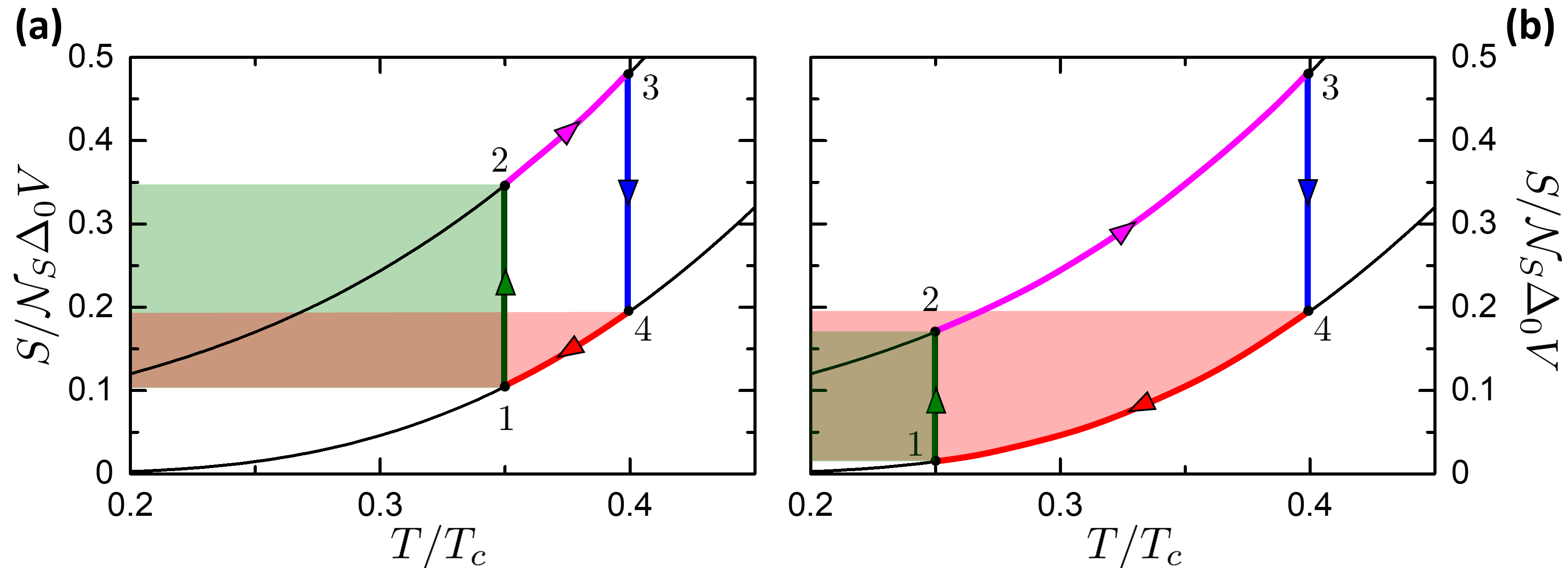}
		\caption{Particular examples of the Josephson-Stirling cycle for $T_R<T_L$. (a) Stirling inverse cycle working as refrigerator. The heat absorbed from the R reservoir in the process $\bf 1 \to 2$, represented by the area defined by the related green arrow, is bigger than the heat released to R reservoir in the process $\bf 4\to 1$, represented by the area defined by the related red arrow. (b) Stirling inverse cycle working as Joule pump, exploiting work to release heat to both reservoirs.} 
		\label{fig:StirlingDegenerate}
	\end{figure}

	In order to work as an engine, it must be $T_R> T_L$, as shown in Figure \ref{fig:StirlingScheme}a. If $T_L > T_R$, the cycle is reversed as displayed in Figure \ref{fig:StirlingDegenerate}. Panels (a) and (b) show the case of $T_R = 0.6T_c$ and $T_L=0.35T_c$ and $T_L=0.25T_c$ respectively. In this case, the machine can work as a refrigerator with the CS represented by the R reservoir and HS represented by the L one (differently from the case of the Josephson-Otto cycle).
	
	Even though the cycles in both panels are clockwise, only the cycle in panel (a) works as refrigerator. Indeed, there are further conditions that define the $(T_L,T_R)$ region where the cycle can work as a refrigerator. Let us consider the heat exchanged with the cold right reservoir, given by processes $\bf 4 \to 1$ and $\bf 1 \to 2$. From Figure \ref{fig:StirlingDegenerate} it can be noticed that in $\bf 4 \to 1$ the heat is released from the system to the R reservoir, while in $\bf 1 \to 2$ the heat is absorbed by the system. Cooling then can take place if $Q_{cs}=Q_R >0$, i.e.,  if $|Q_{12}|>|Q_{41}|$. This is true when $T_L$ is closely below $T_R$, $T_L\lessapprox T_R$; then, when the CS is cooled down, $|Q_{12}|$ decreases, since $\delta S(\ph=\pi,T)$ in an isothermal heat exchange (\ref{eq:IsothermalHeat}) decreases, while $|Q_{41}|$ increases with the increase of the temperature difference $T_{hs}4-T_{cs}$ in the isophasic process. As a consequence, it exists a minimum achievable temperature $T_{MAT}$  that is characterized by a null cooling power $Q_R=0$, i.e., 
	\begin{equation}
	T_{\rm MAT} = T_R {\qquad \rm t.c. \qquad}  Q_{12}(T_L,T_R)+Q_{41}(T_L,T_R)=0 \, \, .
	\end{equation}
	Note that $T_{MAT}$ is a function of the HS temperature $T_L$. If $T_R<T_{MAT}$, the total heat $Q_R$ exchanged with the R reservoir is negative, and the CS is heated. This case corresponds to the $(T,S)$ diagram in Figure \ref{fig:StirlingDegenerate}b, where the red area representing the released heat to the right reservoir includes the green area of the absorbed heat from the right reservoir.
	As before, we call the curve $(T_L,T_R=T_{MAT}(T_L))$ the characteristic curve of the Josephson-Stirling cycle.
	We observe for completeness that when $T_R< T_{MAT}$ and $Q_R<0$, also the left reservoir can absorb or release heat $Q_L=Q_{23}+Q_{34}$, depending on $(T_L,T_R)$. If $Q_L>0$ the cycle absorbs work to transfer heat from the hot to the cold reservoir, constituting a Cold Pump similar to the situation described in the Josephson-Otto cycle. On the other hand, if $Q_L< 0$, the machine releases heat to both the reservoirs, converting completely the work in heat. Following the definition of References \cite{dickerson2016,bizarro2017}, we call this operating mode as {Joule pump}.
	
	In the refrigerator case, the total work is $W=W_{12}+W_{34}$ and the heat extracted is $Q_{cs} = Q_R = Q_{12}+Q_{41}$, like the engine case.
	
\begin{figure}[t]
		\centering
		\includegraphics[width=0.9\textwidth]{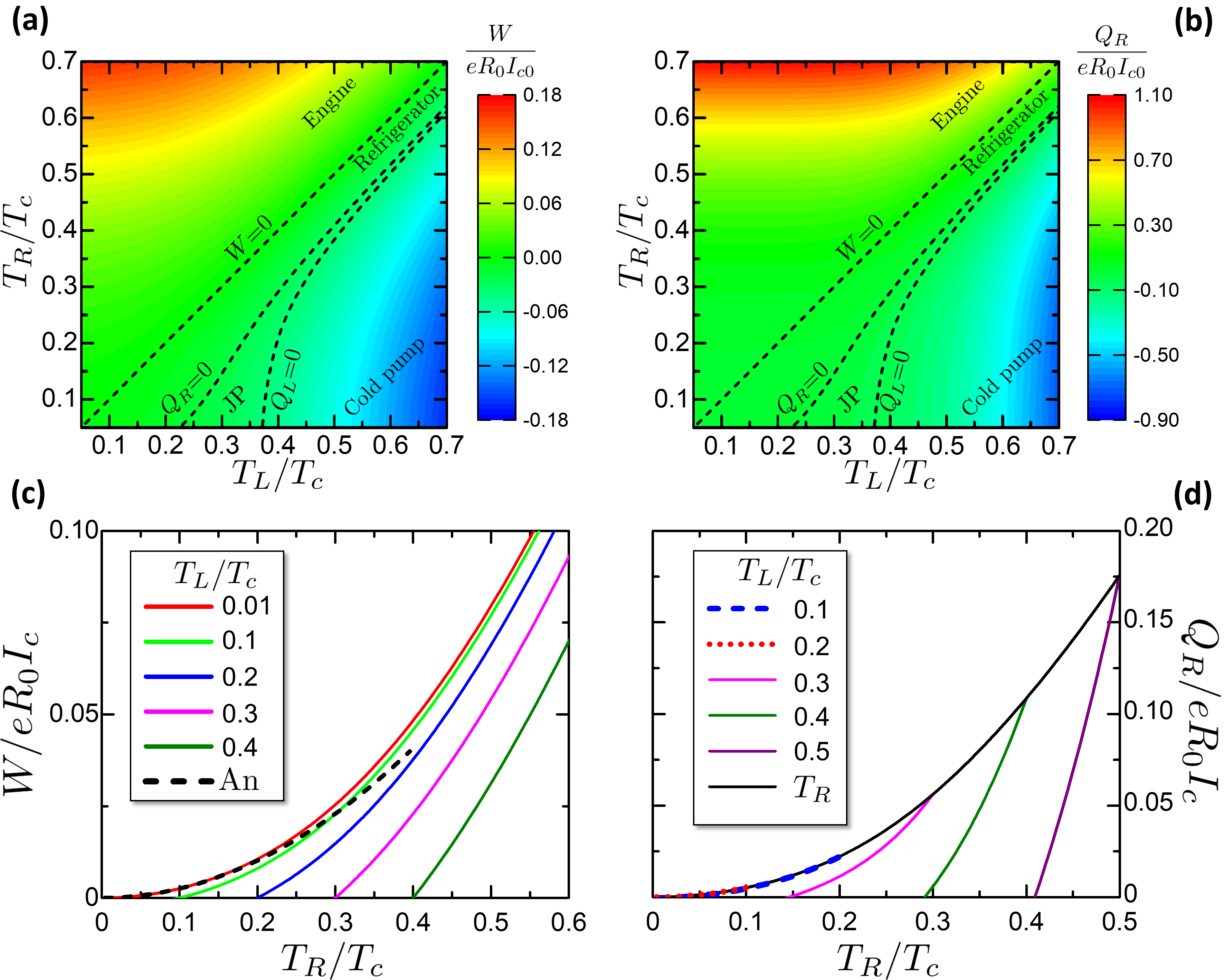}
		\caption{
			({\bf a}) Work released in a Stirling cycle as a function of $(T_L,T_R)$. The dashed curve $W=0$ correspond to the thermal equilibrium curve $T_L=T_R$ and separates the region where the cycle operates as engine or refrigerator. Moreover, the curves $Q_R=0$ and $Q_L=0$ further distinguish regions where the cycle is a Joule Pump (JP) or a Cold Pump. ({\bf b}) Heat absorbed in a Stirling cycle. In both engine and refrigerator modes, the heat $Q_R$ is absorbed from the R reservoir that plays the role of Hot Reservoir or CS in the respective regions. ({\bf c}) Cuts of the work in panel (\textbf{a}) versus the Hot Reservoir temperature $T_R$ for fixed temperatures $T_L$ of the Cold Reservoir. The black dashed line reports expression (\ref{eq:StirlingWAn}). ({\bf d}) Cuts of the absorbed heat $Q_R$ versus the CS temperature $T_R$ for fixed temperatures $T_L$ of the Heat Sink. The black solid curve reports the absorbed heat at $T_L=T_R$. The curves have been obtained with $\alpha=0.6$.
		}
		\label{fig:StirlingWQ}
	\end{figure}

	The released work $W$ and the heat absorbed $Q_{cs}$, $Q_{hr}$ are summarized in Figure \ref{fig:StirlingWQ}. In the color plots in panels a,b, the curves $W=0$, $Q_R=0,Q_L=0$ separate the regions of the engine, the refrigerator, the Joule pump and the cold pump. The curve $W=0$ corresponds to $T_L=T_R$. The~refrigerator region is between the curve $T_L=T_R$ and the characteristic curve $T_{MAT}(T_L)$.
	Figure \ref{fig:StirlingWQ}c reports cuts the released work per cycle versus the HR temperature $T_R$ for fixed CR temperatures $T_L$. The curves reach zero at $T_L=T_R$. The general trend is that the work increases with increasing the temperature difference $T_R-T_L$ between the two reservoirs. 
	
	An analytical expression for $W$ can be calculated. Let us consider a Stirling cycle with $T_L\ll T_R\ll \Delta_0$. The released work can be approximated by $W=\ElEn(\ph=\pi,T_R)-\ElEn(\ph=\pi,T_L\approx0)$. Using approximation (\ref{eq:ReleasedWorkLowT}) for $\ElEn$, we obtain  
	\begin{equation}
	W \approx \frac{\pi}{3\kappa}eR_0 I_c \left(\frac{T_R}{\Delta_0}\right)^2 
	\label{eq:StirlingWAn} \, \, .
	\end{equation}
	This expression is plotted in Figure \ref{fig:StirlingWQ}c and is in good agreement with the numerical results. 
	
	Figure \ref{fig:StirlingWQ}d reports the heat absorbed per cycle $Q_{cs}=Q_R$ versus the CS temperature $T_R$ for fixed HS temperatures $T_L$. The curves go to zero on their left in correspondence of the characteristic curve. The curves are limited on the right by the black curve of $Q_{cs}$ at $T_L=T_R$.  The order of magnitude of the absorbed heat per cycle is $\sim 0.1 e R_0 I_c$.

\begin{figure}[t]
		\centering
		\includegraphics[width=0.95\textwidth]{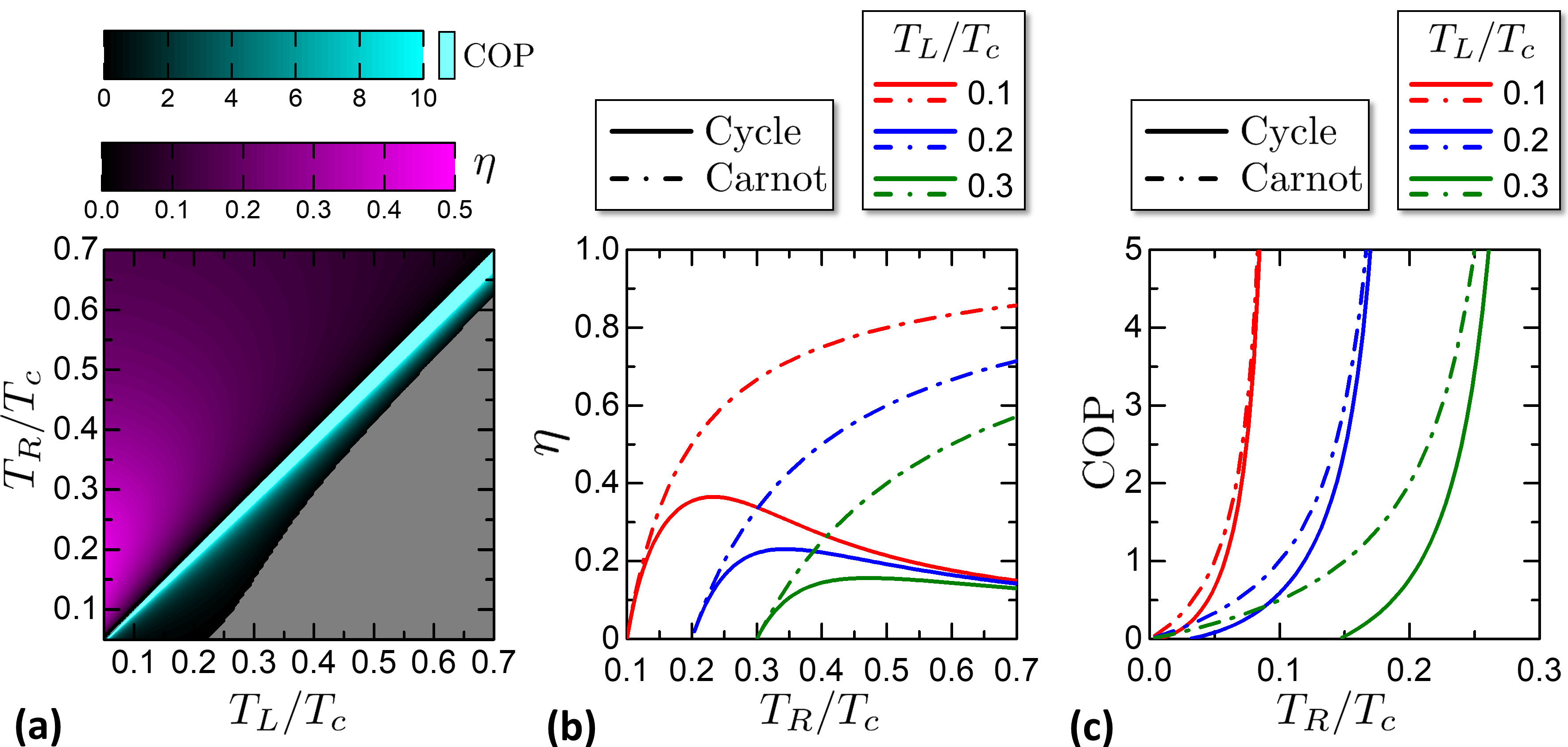}
		\caption{Efficiency and COP of the Stirling machine. ({\bf a}) Color plot of $\eta$ and $\COP$ versus $(T_L,T_R)$, with different color palettes. The gray region represents where the cycle is a Joule pump or Cold pump. ({\bf b}) Cuts of Stirling cycle efficiency $\eta$ versus $T_R$ for chosen $T_L$ in legend. The dot-dashed line reports the Carnot limit to efficiency. The curves end at $T_R=T_L$. ({\bf c}) Cuts of Stirling cycle COP versus $T_R$ for chosen $T_L$ in legend. The dot-dashed line reports the Carnot limit to the COP. The curves go to infinity on the right at the thermal equilibrium state $T_L=T_R$; on the left, the curves are limited by the Stirling characteristic curve.
		}
		\label{fig:StirlingEfficiency}
	\end{figure}

	From the $W,Q_{hr},Q_{cs}$ characteristics it is possible to calculate the $\eta$ and the COP, as reported in Figure \ref{fig:StirlingEfficiency}. Figure \ref{fig:StirlingEfficiency}a shows a color plot of $\eta$ and $\COP$ versus $(T_L,T_R)$. The two quantities are plotted over the engine and refrigerator regions respectively. The gray area is where the cycle works as cold pump or Joule pump. 
	
	Figure \ref{fig:StirlingEfficiency}b reports cuts of the efficiency $\eta$ versus the hot reservoir temperature $T_R$ for chosen ambient cold reservoir temperatures $T_L$. The curves end on the left at the state $T_R=T_L$. For $T_L\to T_R$, both the cycle efficiency and Carnot limit tend to zero. Indeed, the work $W=Q=Q_R+Q_L$ tends to zero, since $Q_R\to - Q_L$ as can be noticed from Figure \ref{fig:StirlingDegenerate}a. 
	
	Figure \ref{fig:StirlingEfficiency}c reports cuts of the COP versus the CS temperature $T_R$ for chosen HS temperatures $T_L$. The curves end on the left at $T_R=T_{MAT}$. For $T_L\to T_R$, both the cycle COP and  its Carnot limit tend to infinity. With the same geometrical argument used for the efficiency, it is $W\to 0 $ and $Q_R = T_R\delta S(\ph=\pi,T)$, implying that the $\COP = Q_R/|W| \to \infty$.
	
	\section{Experimental Feasibility}
	\label{sec:Experimental}
	Here, we briefly comment on some experimental aspects that have to be considered to implement and measure thermodynamic quantities discussed above, {based on hybrid junctions}. The two crucial assumptions of this
	paper are: (i) the processes are quasi-static and (ii) the system is thermally isolated.
	
	{As usually done in the thermodynamics, one is interested to investigate the performance of a thermodynamic cycles in the adiabatic limit (slow evolution) in order to avoid any additional irreversibility due to non-equilibrium processes. Anyway, in any practical realization, one need to develop a cycle in a finite time so it is fundamental to discuss which are the fundamental timescales for the validity of the quasi-static assumption. Hereafter we discuss the quasi-static assumption. It puts a limit on the speed of a process and hence to the cycling frequency $\nu$. In our system, the equilibrium is determined by thermalization of the electron system, and hence the time of equilibration is set by the electron-electron thermal relaxation time $\tau_{e-e}$. Non-equilibrium experiments have been performed in superconducting systems for probing the time-scales of thermal relaxation \cite{kopnin2009,barends2008,gousev1994}. } For aluminium,
	$\tau_{e-e} \approx 1-10\,$ns close to $T_c$ and increases by decreasing the temperature \cite{kopnin2009}. Below $T\lessapprox0.1T_c$ it has been measured that $\tau_{e-e}$
	saturates at $\approx10^2-10^3 \,\SI{}{\micro\second}$ \cite{barends2008}. On the other side, a material with very low $\tau$ can be niobium nitride, where the electron-phonon relaxation
	time is 200 ps \cite{gousev1994}, suggesting a the same order of magnitude for the electron-electron relaxation time. The minimum time interval for a process to be quasi-static can range from the milli-second to
	hundreds of pico-second. {Hence, we~conclude that the rate at which a process or cycle can be performed depends largely on the material and temperature ranges and it is a fundamental issue related to the specific device realization. Even the presence of impurity scattering can alter the relaxation time \cite{sergeev2000}. Experiments suggest that this rate can range from the KHz to tenths of GHz \cite{barends2008,gousev1994,kopnin2009}.  }
	
	The second assumption concerns the thermal insulation. In a superconductor, like in any metal, the electron system is in thermal contact with the phonon
	system. In the superconducting case, the heat flow is exponentially suppressed at low temperatures and scales like the volume $V$ of the \mbox{device \cite{kopninbook,timoveev2009,maisi2013}}.
	Since the electron-phonon thermal conductance is an intrinsic property of the metal and can not be avoided, it is relevant to estimate a threshold $\dot
	Q_{th}$ to the heat leakage below which an isentropic process can be observed. In order to calculate this quantity, let us consider a process that drives
	the phase from $\ph=0$ to $\ph=\pi$ in a time $\tau$. In the ideal case of an isolated system, the process is isentropic. In the real case, a certain amount of
	heat will be absorbed due to the closure of the minigap. If the electron system is well thermalized with the phonon system (corresponding to most of
	real cases), the process is isothermal and absorbs an average heat power
	\begin{equation}
	\dot Q_{th} = \frac{Q}{\tau} = \frac{2\pi}{3 \kappa} \left(\frac{T}{\Delta_0}\right)^2  \frac{e R_0 I_c}{\tau}\, \, 
	\label{eq:conclusion1}
	\end{equation}
	where $Q$ is the heat exchanged during the isothermal process, given by Equation (\ref{eq:IsothermalHeatAn}). As a consequence, isentropic effects can
	be observed if the heat leakage of the electron system allows a power flow that is negligible respect to (\ref{eq:conclusion1}). In particular, for fast
	processes driven at the quasi-static limit, the threshold $\dot Q_{th}$ is maximized to $\dot Q_{th,m}$:
	\begin{equation}
	\dot Q_{th,m} = \frac{Q}{\tau} = \frac{2\pi}{3 \kappa} \left(\frac{T}{\Delta_0}\right)^2  \frac{e R_0 I_c}{\tau_{e-e}}\, \, .
	\label{eq:conclusion2}
	\end{equation}
	If the heat leakage of the system is above this value, an isentropic process can not be observed. At~$T=0.1\Delta_0$ and $\tau\approx10\,$ns it is $\dot
	Q_{th}\approx\SI{3E-8}{\watt \ampere^{-1}} I_c$. For $I_c \approx 1\,$mA, the threshold is $\dot Q_{th}=30\,$pW.
	Based on these considerations, a first experimental setting for testing our findings consists of the measurement of heat
	capacity for different values of phase difference $\ph$. In this case, no external reservoirs neither heat valves are
	required. The measure can be performed by heating up the device with a fixed amount of heat $Q_{test}$ (through a fixed power pulse) and subsequently measure
	the temperature increase $\Delta T$. The quantity $\overline C = Q_{test}/\Delta T$, that approximates the heat capacity $C(\ph,T)$, is~dependent on the
	phase $\ph$ according to the results of Section \ref{sec:Isophasics}. In particular, the relative difference of heat capacity is $\overline C (\ph=\pi,T)/\overline C (\ph=0,T)
	\propto \alpha (T/\Delta_0)^{5/2} e^{T/\Delta_0}$, that is strongly enhanced at low temperatures and is proportional to the parameter $\alpha\propto I_c/V$.
	Interestingly, for these experiments a perfect thermal isolation is not a crucial task, even though it would make the effect more evident. The presence of a certain
	thermal leakage would pull the device to the bath temperature after the power pulse, making the thermometry more difficult. In this case, a more complete
	thermal model is necessary to describe the device behavior.
	
	Another possibility is based on the measurement of temperature variation during an isoentropic process. In this case, no external reservoirs neither heat valves are required,
	but the thermal isolation is a crucial element. As discussed above, the heat leakage in a process $\ph=0\to\pi$ in a time interval $\tau$ must be negligible respect
	to $\dot Q_{th}$ in Equation (\ref{eq:conclusion1}). The heat leakage threshold can be increased by speeding up the process, i.e.,  reducing $\tau$ toward
	the limit $\tau_{e-e}$.
	
	Finally, the most challenging but direct experiment is the cooling of a subsystem with a refrigeration cycle. In this case, external reservoirs, heat valves and the thermal
	isolation of the whole device (ring, valves, heat channels, cooled subsystem) are required. In detail, in order to observe the cooling effect, the spurious
	heat leakage must be lower respect to the cooling power at the thermal equilibrium. As an example, let us consider the Otto cycle removed heat per cycle in
	Equation (\ref{eq:OttoQcsAn}). The~cooling power is $\pi \nu (T/\Delta_0)^2 e R_0 I_c/3\kappa$, where $\nu$ is the cycling frequency. For $\nu=100\,$MHz,
	$T=0.1T_c$, $R_0=\SI{12.9}{\kilo\ohm}$ we have a cooling power of $0.2\,$pW.
	
	{ Before concluding, a comment on the presence of defects at the SN interfaces is in order. Indeed, in SNS systems, the main source of defects is the quality of the SN interfaces. The opacity of the SN contacts can make the proximity effect weaker till disappearance in the tunnel limit \cite{rabani2009,golubov2004}. However, several experiments on quasi-particle tunneling in SNS junctions show that the induced mini-gap has reached good quality features over time \cite{lesueur2008,dambrosio2015}, suggesting the possibility of making good entropy variations in such devices.
		Finally, opaque interfaces imply that the CPR does not follow a KO form, so a different relationship between phase and entropy is expected. However, the Maxwell consistency relation is universal and must hold for every kind of CPR and the entropy variation can be obtained similarly as we proceeded for non-clean interfaces.
	}

	%%%%%%%%%%%%%%%%%%%%%%%%%%%%%%%%%%%%%%%%%%
	\section{Conclusions}
	\label{sec:Discussion}
	In this paper, we have reviewed and analyzed a proximized system with a phase-biased SNS junction under the thermodynamic point of view. By means of arguments of thermodynamic consistency, we have obtained the phase-dependent entropy of the system from its current-phase relation, that we assumed to a Kulik-Omel'yanchuk form. The entropy phase-dependence is related to the presence of an induced minigap in the density of states of the weak link; the minigap depends on the phase $\ph$ across the junction, yielding the phase dependence of the available states and hence of entropy. We obtained closed-form expressions of the entropy for low temperatures and $\ph=0$, $\ph=\pi$. These expressions evidence a strong difference in the temperature dependence for the two phases, where the former is exponentially suppressed and the latter is linear. The entropy relative variation on the phase difference $\ph$ scales like the ratio of the critical current over the system volume. Hence, a~stronger effect requires a higher critical current or smaller volume of the system.
	
	We have discussed equilibrium thermodynamic quantities under quasi-static conditions, obtained by means of the Maxwell equations, investigating processes where phase-coherent properties are linked with thermal properties. In detail, this approach envisions two particular physically observable effects. First, the heat capacity of a proximized system is phase-dependent, passing from an exponentially suppressed behavior at $\ph=0$ to a linear behavior at $\ph=\pi$. Second, the electronic temperature is subject to an isentropic cooling if the system is kept thermally isolated.
	
	Finally, guided by the analogy of the Josephson-base thermodynamics with the classical thermodynamics, we discuss different kind of thermodynamic processes such as isothermal, isophasic and isentropic. After we combine these transformations to define the Josephson-Otto and the Josephson-Stirling cycles, which combine quantum coherence and Josephson effect. This requires the system to be connected to two different reservoirs through heat valves that can allow or stop the flow of heat from them. We characterized the cycle performances in terms of efficiency and COP. The Otto cycle, in particular, shows an interesting capability of having a cooling power till sub-milli-kelvin temperatures.
	
	Further developments can be argued, including Processes that involve a current bias of the junction can be studied. In this case, the phase is a function of the temperature and iso-current processes can be conceived, that can be exploited in different cycles, like Brayton or Diesel cycles.

	\vspace{6pt} 
	
	%%%%%%%%%%%%%%%%%%%%%%%%%%%%%%%%%%%%%%%%%%
	\authorcontributions{Concept and design of the study, F.V., M.C., A.B., P.V. and F.G.; coding and simulations, F.V.; numerical and analytical results analysis, F.V., M.C., A.B., P.V.; discussion and writing, F.V., M.C., A.B., P.V. and F.G.}
	
	%%%%%%%%%%%%%%%%%%%%%%%%%%%%%%%%%%%%%%%%%%
	\funding{The authors acknowledge the European Research Council under the European Unions Seventh Frame-work Programme (FP7/2007-2013)/ERC Grant No. 615187 - COMANCHE, the European Unions Horizon 2020 research and innovation programme under the grant no. 777222 ATTRACT (Project T-CONVERSE), the Horizon research and innovation programme under grant agreement No. 800923 (SUPERTED), the Tuscany Region under the FARFAS 2014 project SCIADRO. M. C. acknowledges support from the Quant-Eranet project ``SuperTop''. A.B. acknowledges the	CNR-CONICET cooperation program Energy conversion in quantum nanoscale hybrid devices, the SNS-WIS joint lab QUANTRA, funded by the Italian Ministry of Foreign Affairs and International Cooperation and the Royal Society through the International Exchanges between the UK and Italy (Grant No. IES R3 170054 and IEC R2 192166).}
	
	%%%%%%%%%%%%%%%%%%%%%%%%%%%%%%%%%%%%%%%%%%
	
	%%%%%%%%%%%%%%%%%%%%%%%%%%%%%%%%%%%%%%%%%%
	\conflictsofinterest{The authors declare no conflict of interest. The funders had no role in the design of the study, in the writing of the manuscript, or in the decision to publish the results.}

	%%%%%%%%%%%%%%%%%%%%%%%%%%%%%%%%%%%%%%%%%%
	%% optional
	\appendixtitles{yes} %Leave argument "no" if all appendix headings stay EMPTY (then no dot is printed after "Appendix A"). If the appendix sections contain a heading then change the argument to "yes".
	\appendix
	\section{Thermodynamics Close to the Critical Temperature}
	\label{sec:Limits}

\begin{figure}[t]
		\centering
		\includegraphics[width=0.95\textwidth]{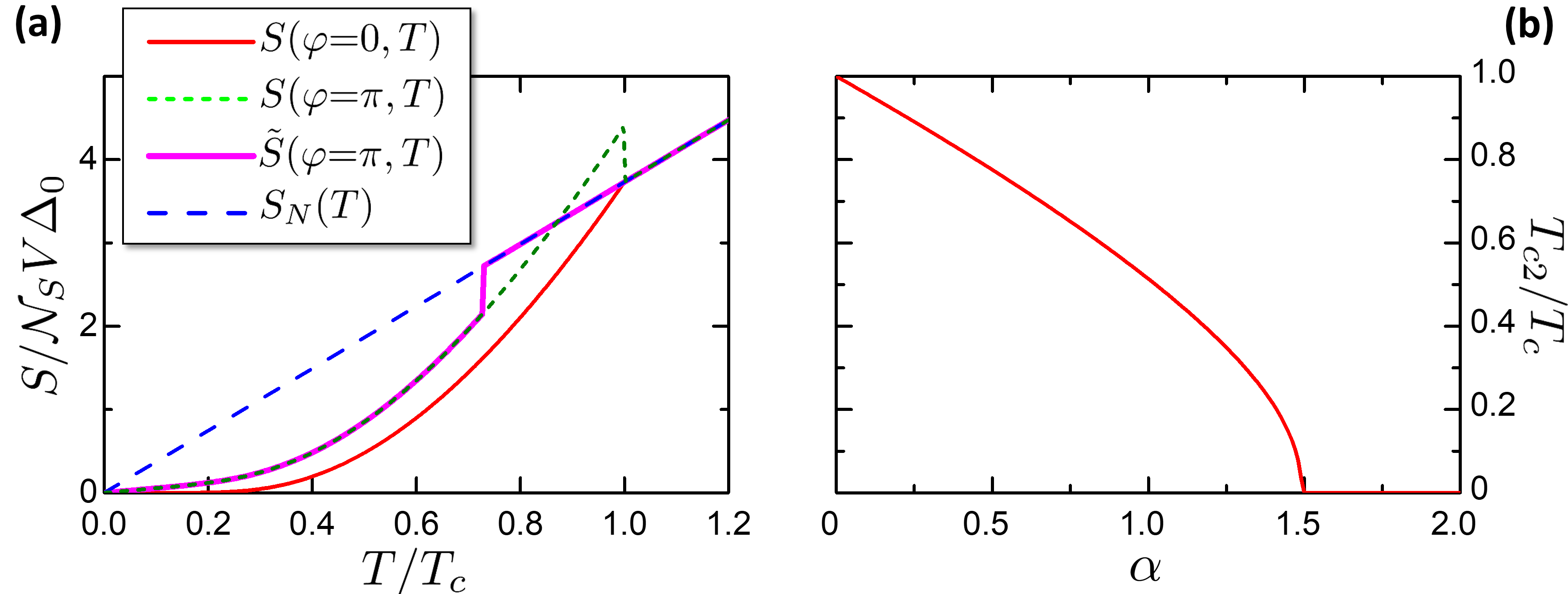}
		\caption{({\bf a}) Entropy scheme close to the critical temperature. $\tilde S$ is equal to $S$ where a phase transition is imposed at $T_{c2}$, calculated with the method in the text. $S_N(T)$ is the normal metal entropy. ({\bf b})~Dependence of the critical temperature of the system $T_{c2}$ at $\ph=\pi$,  normalized to the bulk critical temperature $T_c$, versus $\alpha$.}
		\label{fig:EntropyScheme2}
	\end{figure}

	In this appendix we briefly discuss a particular limit of our theory when the system temperature approaches the critical temperature.
	Apparently, our approach would bring to a thermodynamic inconsistency at temperature close to the critical temperature. Let us indeed consider the entropy scheme in Figure \ref{fig:EntropyScheme2}a. The red solid curve and the dashed green curve report respectively the quantities $S(\ph=0,T)=S_0(T)$ and $S(\ph=\pi,T)=S_0(T)+\delta S(\ph=\pi,T)$. For $T> T_c$, the entropy is given by the normal metal form
	\begin{equation}
	S_N (T) = \frac{2\pi^2 \mathcal{N}_S V T}{3}  \, \, 
	\end{equation}
	plotted over the whole temperature interval as reference (blue dashed line). As can be noticed from Figure \ref{fig:EntropyScheme2}a, $S(\ph=\pi,T)$ has a decreasing jump $\Delta S = S(T\to T_c^+)-S(\ph=\pi,T\to T_c^-)$at $T=T_c$, corresponding to a first-order transition with negative latent heat $T_c \Delta S$. According this scheme, the~system releases heat from the superconducting state  to the normal state, that is unphysical. 
	
	This problem has been discussed and solved in References \cite{vandenbrink1997a,vandenbrink1997b}. The unphysical states close to the transition are metastable and do not correspond to the minimum of the system free energy. The physical stable state is instead given by a null order parameter, that is the normal metal state. The physical solution, instead, consists in a first-order phase transition at a  critical temperature $T_{c2}$, lower than the bulk critical temperature $T_c$, that brings the system from a normal metal state to a superconducting state with absorption of latent heat. An experiment returned results in agreement with this theory \cite{vleeming1997}. 
	
	Following the guidelines of  \cite{vandenbrink1997a,vandenbrink1997b}, we discuss here a simplified macroscopical approach that yields physically acceptable result and allows to grasp the underlying physics.
	
	Let us consider the free energy of both the normal state $F_N(T)$ and superconducting state $F_S(\ph,T)$. At $(\ph=0,T_c)$ the two free energies are equal, $F_S(\ph=0,T)=F_N(T)=F_0$. In their neighborhood on the plane $(\ph,T)$, it is
	\begin{equation}
	F_N(T) = F_0 - \int_{T_c} ^T S_N(T) \dd T = F_0 + \frac{\pi^2 {\cal N}_S V}{3} (T_c^2-T^2) \, \, 
	\end{equation}
	\begin{equation}
	F_S(\ph,T) = F_0 - \int_{T_c} ^T S_0(T) \dd T +\ElEn (\ph,T) \, \, .
	\label{eq:FS}
	\end{equation}
	According the minimization of the free energy, the device is in a superconducting phase in the region $(\ph,T)$ where $F_S(\ph,T)-F_N(T)<0$. The boundary between the two region defines critical temperature $T_{c2}(\ph)$, that is a $2\pi$ periodic in $\ph$ with a minimum in $\ph=\pi$. Over this boundary, the transition is of the first-order (but at $\ph=0$ where takes place the well known second order transition) with a jump in the entropy and associated latent heat. In Figure \ref{fig:EntropyScheme}a, the dashed green line  shows the entropy scheme for this first-order transition at  $\ph=\pi$ for $\alpha=0.6$. In this case, the transition temperature is $T_{c2}(\ph=\pi)\approx0.7T_c$.
	
	The~transition temperature depends on $\alpha$, since it rules the ratio between $S_0$ and $\ElEn$ in (\ref{eq:FS}). The dependence of $T_{c2}(\ph=\pi)$ versus $\alpha$ is reported in Figure \ref{fig:EntropyScheme}b, normalized to the bulk critical temperature. As expected, $T_{c2}(\ph=\pi)\to T_c$ for $\alpha \to 0$. This calculation returns that the superconductivity is suppressed at $\alpha\approx 1.5$, setting practically an upper limit to the supercurrent density.
	
	%%%%%%%%%%%%%%%%%%%%%%%%%%%%%%%%%%%%%%%%%%
	\reftitle{References}
	
	
	%\externalbibliography{yes}
	%\bibliography{D:/DRIVE/PhD_Thesis/Bibliography/ciao.bib}

% The following MDPI journals use author-date citation: Arts, Econometrics, Economies, Genealogy, Humanities, IJFS, JRFM, Laws, Religions, Risks, Social Sciences. For those journals, please follow the formatting guidelines on http://www.mdpi.com/authors/references
% To cite two works by the same author: \citeauthor{ref-journal-1a} (\citeyear{ref-journal-1a}, \citeyear{ref-journal-1b}). This produces: Whittaker (1967, 1975)
% To cite two works by the same author with specific pages: \citeauthor{ref-journal-3a} (\citeyear{ref-journal-3a}, p. 328; \citeyear{ref-journal-3b}, p.475). This produces: Wong (1999, p. 328; 2000, p. 475)

\end{document}